\documentclass[sigconf]{acmart}
\AtBeginDocument{%
  \providecommand\BibTeX{{%
    \normalfont B\kern-0.5em{\scshape i\kern-0.25em b}\kern-0.8em\TeX}}}

\copyrightyear{2023}
\acmYear{2023}
\setcopyright{acmlicensed}\acmConference[WPES '23]{Proceedings of the 21st Workshop on Privacy in the Electronic Society}{November 26, 2023}{Copenhagen, Denmark}
\acmBooktitle{Proceedings of the 21st Workshop on Privacy in the Electronic Society (WPES '23), November 26, 2023, Copenhagen, Denmark}
\acmPrice{15.00}
\acmDOI{10.1145/3603216.3624964}
\acmISBN{979-8-4007-0235-8/23/11}

\usepackage{algorithmic}
\usepackage{graphicx}
\usepackage{textcomp}
\usepackage{xcolor}
\def\BibTeX{{\rm B\kern-.05em{\sc i\kern-.025em b}\kern-.08em
    T\kern-.1667em\lower.7ex\hbox{E}\kern-.125emX}}

\usepackage{enumitem}
\usepackage{amsmath,amsfonts}
\usepackage{algorithmic}
\usepackage{graphicx}
\usepackage{textcomp}
\usepackage{bbm}
\usepackage{tcolorbox}
\usepackage{xcolor}
\usepackage{stmaryrd}
\usepackage[ruled,noline, linesnumbered]{algorithm2e} 
\usepackage{makecell}
\usepackage{xspace}
\usepackage{bm}
\usepackage{caption}
\usepackage{subcaption}
\usepackage{booktabs}
\usepackage{multirow}

\SetAlgoNoLine
\SetAlCapNameFnt{\small}
\SetAlCapFnt{\small}

\newcommand{\mbf}[1]{{\mathbf{#1}}}

\newcommand{\our}{\ensuremath{\mathsf{PROLIN}}\xspace}
\newcommand{\mx}{\mathbf{x}}
\newcommand{\mA}{\mathbf{A}}
\newcommand{\mb}{\mathbf{b}}

\newcommand{\Tgd}{T_{\mathsf{gd}}}

\newcommand{\argmin}{\mathop{\mathrm{argmin}}\limits} 
\newcommand{\argmax}{\mathop{\mathrm{argmax}}\limits}

\begin{document}

\title[Client-specific Property Inference against Secure Aggregation in
Federated Learning]{Client-specific Property Inference against Secure Aggregation in
Federated Learning}

\author{Raouf Kerkouche}
\email{raouf.kerkouche@cispa.de}
\affiliation{%
  \institution{CISPA Helmholtz Center for Information Security}
  \country{Germany}
}

\author{Gergely \'Acs}
\email{acs@crysys.hu}
\affiliation{%
  \institution{Department of Networked Systems and Services, CrySyS Lab, BME}
  \country{Hungary}
}

\author{Mario Fritz}
\email{fritz@cispa.de}
\affiliation{%
  \institution{CISPA Helmholtz Center for Information Security}
  \country{Germany}
}


\begin{abstract}
Federated learning has become a widely used paradigm for collaboratively training a common model among different participants with the help of a central server that coordinates the training. Although only the model parameters or other model updates are exchanged during the federated training instead of the participant's data, many attacks have shown that it is still possible to infer sensitive information or to reconstruct participant data.
Although differential privacy is considered an effective solution to protect against privacy attacks, it is also criticized for its negative effect on utility. Another possible defense is to use secure aggregation, which allows the server to only access the aggregated update instead of each individual one, and it is often more appealing because it does not degrade the model quality. However, combining only the aggregated updates, which are generated by a different composition of clients in every round, may still allow the inference of some client-specific information.

In this paper, we show that simple linear models can effectively capture client-specific properties only from the aggregated model updates due to the linearity of aggregation. We formulate an optimization problem across different rounds in order to infer a tested property of every client from the output of the linear models, for example, whether they have a specific sample in their training data (membership inference) or whether they misbehave and attempt to degrade the performance of the common model by poisoning attacks. Our reconstruction technique is completely passive and undetectable.
We demonstrate the efficacy of our approach on several scenarios, showing that secure aggregation provides very limited privacy guarantees in practice. The source
code is available at \href{https://github.com/raouf-kerkouche/PROLIN}{\color{magenta}{https://github.com/raouf-kerkouche/PROLIN.}} 
\end{abstract}





\maketitle

\section{Introduction}

Machine learning models have made their way into a broad range of application domains. However,  accuracy of such models is typically dependent on the amount of available data. Often a lack of sufficient data prevents training of accurate models. One of the simplest solutions is collaboration between data holders by sharing data in order to train better models. Yet,  this solution may not be viable if the data in question is sensitive and privacy is crucial.
Federated Learning addresses the above constraints by allowing collaborative training of a model without sharing any data. Instead, only the model parameters are shared between a central server and the different entities that participate in the learning process. Federated learning has become a veritable paradigm and is used to train shared models for many applications, such as input text prediction\cite{GoogleKeyboardPred}, ad selection~\cite{adselection}, drug discovery\cite{melloddy} or various medical applications~\cite{fets,choudhury2019differential,RKerkoucheIHP} that use the confidential data of many different entities.

Unfortunately, even though private training data is not shared directly in federated learning, many attacks have shown that it is possible to infer sensitive information about the training data of each client. Membership attacks \cite{nasr2019comprehensive,Property} allow, for example, to infer whether a specific record is included in a participant's dataset. Similarly, the attack in \cite{Property} allows inferring whether a group of people with a specific property independent of the main task is included in any participant’s dataset. Even worse, it is possible to reconstruct the training data~\cite{DLG,IDLG,VFLllgusenix,LLG,li2021label,NEURIPS2020_c4ede56b,Property,li2022auditing}. 

Solutions exist to remedy the above attacks, such as Differential Privacy. 
Although this can provide a strong privacy guarantee, it can also jeopardize the benefits of federated learning by severely deteriorating the accuracy of the commonly trained model. Hence, many companies are still reluctant to use Differential Privacy, especially in scenarios, where only a limited number of companies engage in training (process) and want to prevent the leakage of any, not only sample-specific information about their abundant training data\footnote{\url{https://www.melloddy.eu}}. However, the small number of clients is usually insufficient to counterbalance the negative effect of noise on model accuracy, which can eventually incur a (business) risk for the clients. 


Secure aggregation \cite{BonawitzIKMMPRS17} is often used as an alternative (or complementary) mitigation technique against unintended information leakage.  This cryptographic solution allows the protocol participants to access only the aggregated model updates but not the individual update of any client sent for aggregation. Indeed, most existing inference and reconstruction attacks rely on accessing the individual gradients (model updates) in order to succeed.   Secure aggregation guarantees that even if any participant learns some confidential information from the aggregated model, they are still unlikely to attribute this information to any specific client without the necessary background knowledge \cite{saimpo}. Albeit providing strictly weaker confidentiality guarantees than differential privacy, secure aggregation does not degrade model accuracy, has small computational overhead, and has therefore become an indispensable part of any federated learning protocol. Although there exist active attacks~\cite{fishing,pasquini,fowl2021robbing,boenisch2021curious,boenisch2023federated} even against secure aggregation, which enforce information leakage by model or data poisoning, these attacks are either detectable, thereby providing evidence of the misdeed, or can be prevented \cite{xu2019verifynet,guo2020v,zhang2020privacy,fu2020vfl,mou2021verifiable,han2022verifiable,jiang2021pflm,madi2021secure,hahn2021versa}. This makes such an active attack less likely in practice, especially if clients can suffer a reputation loss due to the potential repercussions that can easily outbalance the benefit of a successful attack. 

In this paper, we show that secure aggregation often fails to prevent the attribution of confidential information to a client, even if the adversary is only a \emph{passive} observer who faithfully follows the federated learning protocol. Our attribution technique shows that the server or a client who can access only the common model in each training round can learn accurate client-specific information (i.e., a property of the client, such as whether its training data includes some specific samples) without being detected, even if secure aggregation is employed. Our technique 
does not need any background knowledge about any specific client to succeed, just the aggregated common model observed per round, and the identity of the clients participating per round.

We exploit the fact that the composition of participating clients changes in almost every round to decrease communication costs and guarantee convergence. This optimization allows us to solve 
a system of linear equations, where the unknowns are some (private) contributions of the clients whose sums are observable. 
Prior work \cite{lam2021gradient} has shown that if these contributions are the gradients, then simple linear regression can be used to reconstruct the mean gradient vector of every client as long as the variance of the gradient is small per client
and there are a sufficient number of rounds (equations). 
We show that, instead of disaggregating the sum of gradients and then launching a supervised inference attack on the reconstructed individual gradients per client, it is more effective to directly reconstruct the linear features used by this inference attack. In particular, we substantially improve on \cite{lam2021gradient} by leveraging the linearity of model aggregation:
the unknowns are the linear features of an individual model update  that effectively capture property information and are reconstructed from the observed model aggregates.  The (private) property value of every client is computed by maximizing the likelihood of these reconstructed features over the rounds given their prior distributions on the auxiliary dataset. 
Since only a small number of features are reconstructed and used for inference instead of the potentially large gradient vectors, our approach has a significantly smaller variance compared to \cite{lam2021gradient} at the cost of a slight bias. 
Our approach is general and can be utilized to infer various client-specific properties only from the observable aggregations of model updates. Moreover, it is completely passive, unintrusive, and does not intervene in the normal operation of federated learning.

\begin{figure}[ht]
\centering
\includegraphics[scale=0.31]{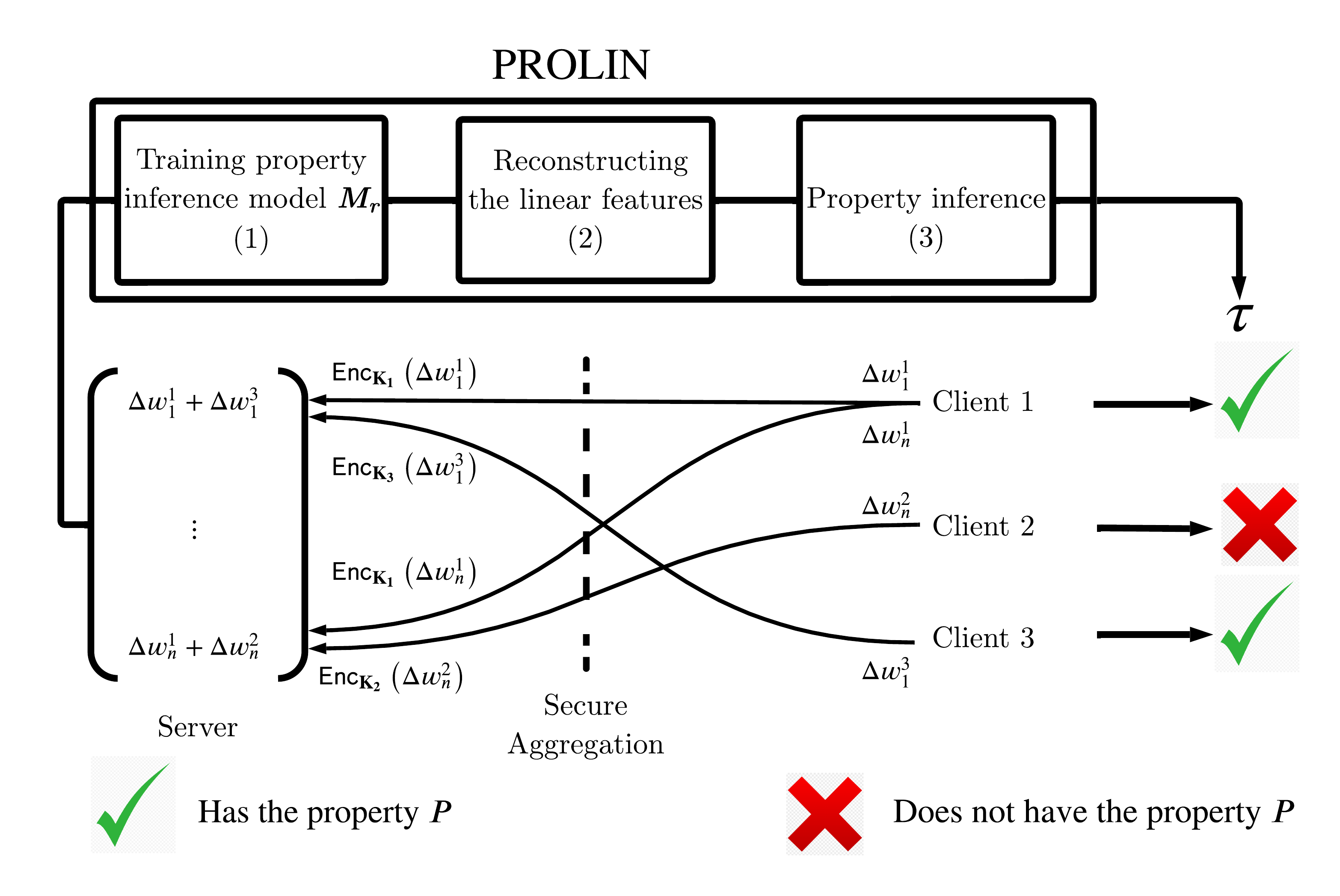} 
\caption{Illustration of \our.}
\label{fig:prolin}
\end{figure}

Our main contributions are the following:
\begin{itemize}[noitemsep,topsep=0pt]
    \item We show that secure aggregation is not sufficient to prevent the reconstruction of client-specific information.
    We propose a general, completely passive reconstruction technique called \our,
    which, exploiting the linearity of model aggregation, 
    uses linear models to capture property information from aggregated model updates and attributes them to specific clients. We demonstrate our general approach on two detection tasks. 
    
    \item   We identify clients whose training data includes a specific target sample. 
     We disaggregate the linear features used by the membership inference attack that yields increased attack accuracy compared to related work  \cite{lam2021gradient}. 
    This negative result shows that accurate private information leakage is still possible with secure aggregation, even without client-specific background knowledge, and that membership inference attacks remain a significant risk, even with a passive adversary.    
    
    \item We detect clients that exhibit malicious behavior by launching (untargeted) poisoning attacks. To the best of our knowledge, prior works have only addressed poisoning detection \emph{without secure aggregation} \cite{FungYB20, RiegerNMS22,NguyenRCYMFMMMZ22,CaoJZG23}.
    This positive result shows that secure aggregation is not enough to hide poisoning attacks, which decreases the incentive and therefore the risk of such attacks.

\end{itemize}
The operation of \our is illustrated in Figure \ref{fig:prolin}.

\section{Background}

\label{sec:backg}
\subsection{Federated Learning}
\label{FL-STANDARD}


In federated learning \cite{ShokriS15,FedAVG}, multiple clients build a common machine learning model from the union of their training data without sharing the data with each other. 
At each round of the training, in order to reduce communication costs, only a fraction $C$ of all $N$ clients are randomly selected to retrieve the common model from the parameter server,  update the global model based on their own training data, and send back their updated model to the server. 
The server aggregates the updated models of all clients to obtain a global model that is re-distributed to some selected parties in the next round.  Different aggregation techniques \cite{sannara2021federated,yurochkin2019bayesian,cho2020client,FedAVG} have been proposed, we consider federated averaging (FedAvg) \cite{FedAVG} in this paper.

More specifically, let $\mathbf{A} \in \{0,1\}^{n\times N}$ denote the participation matrix, where $\mathbf{A}_{r,i} = 1$ if client $i$ participates in round $r$, and 0 otherwise. As the total number of participants is the same in each round, it holds that $\sum_{i} \mathbf{A}_{r, i} = C\cdot N$ for all $r$. 
At round $r$, a participating client $i$ (i.e., $\mathbf{A}_{r,i} = 1$) executes $\Tgd$ local gradient descent iterations on the common model $T_{r-1}$ with parameters $\mbf{w}_{r-1}$, using its own training data $D_i$, and sends the model update $\Delta \mbf{w}_{r}^i = \mbf{w}_{r}^i - \mbf{w}_{r-1}^i$ to the server, which then obtains the new common model $T_r$ by aggregating the received updates as
$\mbf{w}_{r} = \mbf{w}_{r-1} + \sum_{i} \frac{|D_i|}{\sum_j |D_j|} \mathbf{A}_{r,i} \Delta \mbf{w}_{r}^i$ (a client's update is weighted with the size of its training data), where  $z$  denotes the model (update) size \cite{FedAVG}.
Finally, the server re-distributes $\mbf{w}_{r}$ to the clients selected in the next round. The server stops training after a fixed number of rounds $n$ or when the performance of the common model does not improve on held-out data. 


Federated learning is often combined with secure aggregation to prevent the server and any client from accessing the individual updates $\Delta\mathbf{w}_r^i$ rather than just their aggregation $\sum_i \frac{|D_i|}{\sum_j |D_j|} \mathbf{A}_{r,i} \Delta\mathbf{w}_r^i$ per round \cite{AcsC11,bonawitz2016practical}.
When secure aggregation is used, each client encrypts its individual update before sending it to the server. Upon reception, the server sums the encrypted updates as:
\begin{align}
  \sum_{i\in \{j : \mathbf{A}_{r,j}=1\}}\mathsf{Enc}_{\mbf{K}_i}\left( \frac{|D_i|}{\sum_j |D_j|}\Delta \mbf{w}_r^i\right) = \sum_{i}  \frac{|D_i|}{\sum_j |D_j|} \mathbf{A}_{r,i} \Delta \mbf{w}_r^i 
 \label{eq:sec_agg}
\end{align}
where $\mathsf{Enc}_{\mbf{K}_i}( \mathbf{x})= \mathbf{x} + \mbf{K}_i \mod p$ and $\sum_{i\in \{j : \mathbf{A}_{r,j}=1\}}\mbf{K}_i=0$ (see \cite{AcsC11,bonawitz2016practical} for details). Here the modulo is taken element-wise and $p=2^{\lceil \log_{2}(\max_{k}|| \Delta \mbf{w}_r^k||_{\infty}|\mathbb{K}|)\rceil}$.

\subsection{Linear regression}
\label{sec:lin_reg}
Given a linear model as
\begin{align}
\label{eq:linear_model_orig}
\mathbf{b} = \mathbf{A}\mathbf{\hat{x}} + \bm{\delta}
\end{align}
where $\mA \in \mathbb{R}^{n \times N}$ is a known matrix, $\mb \in \mathbb{R}^n$ are the observed (noisy) aggregates, and 
$\bm{\delta} \in \mathbb{R}^{N}$ are random variables describing the noise with zero mean and finite variance.
In machine learning parlance, each row of $\mA$ corresponds to a training sample with $N$ input variables (features), and $\mathbf{\hat{x}}$ is the unobserved parameter vector of the linear model  to be determined.
Eq.~\eqref{eq:linear_model_orig} defines a system of $n$ linear equations for $\hat{\mathbf{x}}$ as unknowns, and the method of ordinary least squares (OLS) provides an unbiased estimate $\mathbf{\widetilde{x}}$ of $\mathbf{\hat{x}}$ as
\begin{align}
\label{eq:ols}
\mathbf{\widetilde{x}} = \argmin_{\mx}(\mb - \mA \mx)^2
\end{align}
that is, $E[\widetilde{\mathbf{x}}] = \mathbf{\hat{x}}$ regardless of $\mA$.
Eq.~\eqref{eq:ols} has the closed-form solution of $\mathbf{\widetilde{x}} = \mathbf{A}^+ \mb$, where $\mA^{+}$ is the Moore-Penrose inverse of $\mA$. 
According to the Gauss-Markov theorem, $\mathbf{\widetilde{x}}$ has the smallest variance among all unbiased estimators 
if $\bm{\delta}_i$ are uncorrelated, have zero mean, and equal variance.
Although  $\mathbf{\widetilde{x}}$ is an unbiased estimate, there are other estimators that exploit the bias-variance trade-off and decrease the variance of the estimate at the cost of introducing some bias by regularization, so that the total error (the sum of squared bias and variance) is still smaller than for any unbiased estimator, including OLS. 

Ridge Regression (RR) provides an $L_2$ regularized estimation of $\mathbf{\hat{x}}$ as
\begin{align}
\label{eq:ridge}
\mathbf{\widetilde{x}}' = \argmin_{\mx}[(\mb - \mA \mx)^2 + \lambda \cdot \mx^\top \mx]
\end{align}
where $\lambda$ is the regularization  parameter. 
RR introduces bias by constraining the set of feasible solutions of the least square problem into a zero-centered $L_2$ ball even if the  real solution $\mathbf{\hat{x}}$ is outside this ball. Compared to OLS, this can significantly reduce the variance and hence the mean squared error of the final estimate $\mathbf{\widetilde{x}}'$, which is especially useful when the variance of $\bm{\delta}$ is too large (e.g., when the number of observations $n$ is too small or the observations are too noisy).
In general, the larger the variance of $\bm{\delta}$, the larger the regularization should be since increased $\lambda$ causes the variance to vanish and the bias will dominate the total estimation error.
Ultimately, the optimal choice of $\lambda$  depends on the distribution of $\mathbf{\hat{x}}$ and $\bm{\delta}$.


\section{Related Work}

\noindent \textbf{Privacy attacks in Federated Learning:}
Several privacy attacks  have been proposed to learn confidential information about the client's training data in federated learning \cite{DLG,IDLG,VFLllgusenix,LLG,li2021label,NEURIPS2020_c4ede56b,Property,li2022auditing, boenisch2021curious, nasr2019comprehensive, lam2021gradient}. 

In \cite{nasr2019comprehensive},  membership inference attack (MIA) is proposed to infer if a specific record is included in the training dataset of the participants. At each round, the adversary first extracts a set of features from every snapshot of the trained global model received from each selected client, such as the output value of the last layers and the hidden layers, the loss values, and the gradient of the loss with respect to the parameters of each layer. These features are used to train a single membership inference model, which is a convolutional neural network, at the end of the training. The attack requires access to each individual update and is therefore ineffective when secure aggregation is used. Finally, the paper has also shown that the attack can be much more effective if the adversary is active instead of passive.  \cite{Property} introduced the first membership attack under federated learning settings that consists of exploiting the non-zero values of the embedding layer and is therefore only valid for this specific type of layer. Moreover, it requires access to individual updates and is thus also ineffective when secure aggregation is used. 

In \cite{lam2021gradient}, the authors reconstruct the participation matrix and then the average update per client. The reconstruction of the participation matrix is out of scope in our paper because, in federated learning, the server selects the participating clients according to their availability in each round and therefore knows the participation matrix. However, the reconstruction of the average update per client is naturally the baseline we will consider in our paper (see Section \ref{sec:baseline}). To the best of our knowledge, \cite{lam2021gradient} is the first and only work that performs disaggregation in federated learning against secure aggregation by considering a passive adversary (the server). After reconstructing the average update per client using ordinary least squares (OLS), the server infers the membership information from these reconstructed updates. We show that disaggregating some linear features of the update vector used by the attacker/detector model (e.g., membership inference model) provides more accurate membership inference than disaggregating the whole update vector. Instead of training a single membership model at the very end of the training, we train a distinct membership inference model in each round and combine their inner representations (features) into a final decision with optimization. This approach is more robust especially if some rounds have very inaccurate inference models. Moreover, we also demonstrate the efficacy of our approach on identifying malicious clients.


Recently, a new line of research has focused on active privacy attacks~\cite{fishing,pasquini,fowl2021robbing,boenisch2021curious,boenisch2023federated}, where a malicious server poisons the parameters of the global model in order to reveal a client's update vector. These attacks try to increase the norm of the update vector for a targeted client while decreasing it for non-targeted clients. Some attacks are more restricted than our proposal because they either require large linear layers after the input layer \cite{fowl2021robbing,boenisch2021curious} or are designed and evaluated only for FedSGD~\cite{FedAVG}, where each client performs a single SGD update~\cite{fishing,pasquini,fowl2021robbing}, unlike FedAvg~\cite{FedAVG}. 
In addition, except for \cite{pasquini}, these active attacks only link the recovered update vector to the set of participating clients in a round and do not combine the recovered updates across rounds to infer a property of a client.
Although a stealthier active attack has been proposed in  \cite{pasquini} that is harder to detect, it is not undetectable, in contrast to passive attacks. In fact, at the cost of additional computational overhead but without harming model quality, all active attacks can be prevented by using cryptographic protocols to verify whether the server manipulates the common model 
\cite{xu2019verifynet,guo2020v,zhang2020privacy,fu2020vfl,mou2021verifiable,han2022verifiable,jiang2021pflm,madi2021secure,hahn2021versa, ijcai2022p792}.

Since active attacks can be detected (or prevented) without degrading model accuracy, they are less practical than passive attacks. Moreover, passive attacks can even be launched offline on more powerful hardware after capturing the protocol messages. \\ 

\noindent \textbf{Poisoning attacks in Federated Learning:}
We focus on integrity attacks \cite{papernot2018sok} and more specifically on poisoning attacks and their defenses.
Poisoning attacks are performed either by manipulating the training data (data poisoning)~\cite{pois_svm,rubinstein2009stealthy,machine_teaching,xiao2015feature,koh2017understanding,chen2017targeted,jagielski2018manipulating,shen2016auror,fung2018mitigating,tolpegin2020data} or by directly manipulating the model update  (model poisoning)~\cite{EPFL_attack,SIGNSGD_vote_robustness,littleisenough,nasr2019comprehensive,kerkouche2020federated} .
These attacks can be either targeted by aiming only at reducing the accuracy of the model on some target classes ~\cite{Backdoor_Fed,IBM_attack} or untargeted, in which case they aim at reducing the accuracy of the model globally without any distinction between the classes~\cite{EPFL_attack,SIGNSGD_vote_robustness,littleisenough,nasr2019comprehensive,kerkouche2020federated}. 

Numerous defenses exist against these attacks, which generally choose the best update in each round~\cite{EPFL_attack,bulyan,tolpegin2020data,fung2018mitigating,chang2019cronus,wang2019neural} or derive a more robust update in each round~\cite{trimmed_mean_median,xie2018generalized,shen2016auror} based, for example, on the median value calculated from the updates sent by the participants to the server~\cite{trimmed_mean_median}. However, they generally require access to each individual update and therefore cannot be employed with secure aggregation because the latter only allows access to the sum of the individual updates. To the best of our knowledge, only \cite{SIGNSGD_vote_robustness} and \cite{kerkouche2020federated} use a more robust update with secure aggregation, however, they also require that each client sends only the sign of each coordinate's value of the update vector, which slows down convergence. Some works also aim to detect clients launching poisoning attacks \cite{FungYB20, RiegerNMS22,NguyenRCYMFMMMZ22,CaoJZG23}, assuming that the model updates of every client  are available for detection in every round.

In our paper, we identify participants with malicious behavior in federated training even if secure aggregation is used. Specifically, we consider two untargeted poisoning attacks called gradient inversion~\cite{SIGNSGD_vote_robustness} and gradient ascent attacks~\cite{nasr2019comprehensive}, which modify the update vector locally so that the performance of the common model declines.

\section{Threat model}
\label{sec:threat_model}
The server can infer two types of properties of \emph{each} client: the occurrence of a given target sample in the client's training data (\emph{membership detection}) and whether the client executes poisoning attacks (\emph{misbehaving detection}). 
In membership detection, the server is  a \emph{semi-honest} adversary who aims to identify all clients that have the target sample in their training data.
In misbehaving detection, the server is a \emph{honest detector} who aims to identify all \emph{malicious clients} that perform a poisoning attack to degrade the performance of the federated model (at most a fraction $\phi$ of all clients are malicious). 
As opposed to previous works \cite{fishing,pasquini,fowl2021robbing,boenisch2021curious,boenisch2023federated}, the server  is  \emph{passive} in both cases, that is, it faithfully follows the federated protocol in Section \ref{FL-STANDARD}. This can be enforced by applying verifiable federated learning schemes \cite{xu2019verifynet,guo2020v,zhang2020privacy,fu2020vfl,mou2021verifiable,han2022verifiable,jiang2021pflm,madi2021secure,hahn2021versa}.


 In misbehaving detection, malicious clients perform poisoning by executing gradient ascent or inversion attacks.
In a \emph{Gradient Ascent Attack} \cite{nasr2019comprehensive}, 
malicious clients aim at maximizing the loss by performing gradient \emph{ascent} instead of descent on their own training data. In particular, they update the model parameters locally as 
$
\mbf{w}_r^i = \mbf{w}_{r-1}^i + \eta \nabla \ell(D_i;\mbf{w}_{r-1}^i)
$,
where $\eta$ is the learning rate and $\ell$ is the loss function. 
This attack attempts to maximize the \emph{average} misclassification rate of the global model and is more effective if the training data of the malicious and benign nodes come from similar distributions.
In a \emph{Gradient Inversion Attack} \cite{SIGNSGD_vote_robustness},
malicious clients faithfully compute their model update $\Delta \mbf{w}_r^i$ but send  $-\Delta \mbf{w}_r^i$ (instead of $\Delta \mbf{w}_r^i$) for aggregation.

Since the model update $\Delta \mbf{w}_r^i$ is computed on the entire local data of a client, the target sample always influences $\Delta \mbf{w}_r^i$ in membership inference if it is included in the training data. Likewise, poisoning is often executed in each round by every malicious client and therefore has a direct impact on $\Delta \mbf{w}_r^i$.
Hence, the server can train a (supervised) binary detector model $M_r$ to recognize such changes in $\Delta \mbf{w}_r^i$ and tell only from the model update of a client whether it has the tested property in round $r$: 
$M_r(\Delta \mathbf{w}_r^i)$ denotes the confidence of the server that the client has the target sample in its training data in membership detection or that it performs poisoning in misbehaving detection. To train the detector model $M_r$, an auxiliary (or shadow) dataset $D^{aux}$ is also available to the server, which has sufficiently similar distribution as the clients' training data, though $D^{aux}$ does not include any training samples of any honest client. The availability of an auxiliary dataset is a natural assumption of any supervised inference model and not specific to our proposal. Our approach can also be generalized to any unsupervised or semi-supervised inference model $M_r$ as long as it uses a linear map of the gradients (see Section \ref{sec:proposal} for details). Also, $D^{aux}$ can be generated synthetically: at the end of the federated learning protocol, the final common model is inverted\footnote{Model inversion can be performed by training a Generative Adversarial Network (GAN) where the discriminator is the final common model and the trained generative model is used to produce synthetic data \cite{TruongMWP21}.} to generate synthetic training data, and $M_r$ is trained with such synthetic data for property reconstruction\footnote{In that case, reconstruction  is performed after federated learning, if all model updates are recorded during training.}. 

For detection, the individual  model updates $\Delta\mathbf{w}_r^i$ are not accessible due to secure aggregation, however,  the server can access and record their sum $\sum_i \mathbf{A}_{r,i} \Delta\mathbf{w}_r^i$ in each round as well as the intermediate snapshots $T_r$ of the common model.
In addition, the complete participation matrix $\mA$ is known to the server,  which is a reasonable assumption. Otherwise, the server can exploit side information to reconstruct $\mA$ (see \cite{lam2021gradient}  for details).

\section{Property reconstruction}
We show how the server can reconstruct the property information of every client accessing only the aggregated model updates. 
We present two reconstruction approaches in this section. In the first naive approach, described in Section \ref{sec:baseline}, the server
disaggregates the sum of update vectors into the individual update of every client and applies the trained detector model $M_r$ on each disaggregated update vector separately.
However, the error of this approach can be proportional to the update (model) size in the worst case. Hence, we improve this naive approach and rather disaggregate the linear features of the aggregated update vector, which are used by the detector model $M_r$. The server finds client-specific properties that maximize the observation probability of these disaggregated features. 
This improved approach is called \our and described in Section \ref{sec:proposal}.

\subsection{Naive property reconstruction with gradient disaggregation}
\label{sec:baseline}

The naive reconstruction technique is based on  \cite{lam2021gradient} and consists of three steps: (1) reconstructing the expected update vector for every client, (2) 
training the detector model $M_r$ per round to predict property $P$ from the reconstructed updates, and (3) combining the per-round model predictions to make the final decision about the property of  each client. \medskip 

\subsubsection{Gradient reconstruction:} 
\label{sec:gradient_recon}
The update vector $\Delta\mathbf{w}_r^i$ of every client $i$ is changing over the rounds due to the stochasticity of learning. Still, it is possible to approximate the mean of these per-round updates of a client (i.e., a single "average" update vector per client) with linear regression as follows.

Suppose that
the aggregation is  described as
\begin{align}
\label{eq:lin_grad}
\mb_r = \sum_i ^N \mathbf{A}_{r,i}\Delta\mathbf{w}_r^i = \sum_{i}^N \mathbf{A}_{r,i} \mathbf{\Delta \hat{w}}_i + \bm{\xi}_r
\end{align}
where $\Delta \mathbf{\hat{w}}_i$ is the expected update vector of client $i$ that we want to reconstruct, and $\bm{\xi}_r$ represents a vector of independent, unobserved random variables that accounts for the aforementioned stochasticity of learning and models the variance of the individual updates over the rounds ($\Delta \mathbf{w}_r^i, \Delta \mathbf{\hat{w}}_i, \bm{\xi}_r \in \mathbb{R}^z$ for all $i$ and $r$).
 Given $\mb_r$ and $\mathbf{A}$, Eq.~\eqref{eq:grad_recon} defines $z$ systems of linear equations (one per update coordinate), each with $n$ equations over $z\times N$ unknowns altogether, which can be approximated by OLS. 
Formally,
\begin{align}
\label{eq:grad_recon}
\mathbf{\widetilde{W}} = \argmin_{\mathbf{x} \in \mathbb{R}^{N \times z}} ||\mathbf{B} - \mathbf{A} \mathbf{x}||_F^2
\end{align}
where $\mathbf{B} = (\mathbf{b}_1, \ldots, \mathbf{b}_n) \in \mathbb{R}^{n \times z}$, and $||\cdot||_F$ is the Frobenius norm.
According to the Gauss-Markov Theorem, $\mathbf{\widetilde{W}}$ is the best unbiased estimator of $\mathbf{\hat{W}} = (\Delta \mathbf{\hat{w}}_1, \ldots, \Delta \mathbf{\hat{w}}_N) \in \mathbb{R}^{N \times z}$ if $\bm{\omega}_1, \ldots, \bm{\omega}_n$ are uncorrelated, have zero mean and identical finite  variance. 


\medskip 

\subsubsection{Training the detector model $M_r$:}
\label{sec:train_Mr_naive}
The server trains a per-round detector model $M_r$  on $D^{aux}$ in order to infer the property $P$ from the reconstructed expected update $\Delta \mathbf{\widetilde{w}}_i$ for each client as follows: 

First, the server creates two disjoint sets of batches $\mathbb{B}^+$ and $\mathbb{B}^-$ from $D^{\mathit{aux}}$, which are used to generate updates with and without property $P$, respectively.
For membership detection, every batch in $\mathbb{B}^+$ includes the target sample whose membership is detected, while every batch in $\mathbb{B}^-$ excludes the same target sample. 
Then, provided with the common model $T_{r-1}$ in round $r$, the server creates the (balanced) training data $D' = D^+ \cup D^-$ such that $D^+ = \{(\Delta \mathbf{w}_r^B, \mathsf{True}) | B \in \mathbb{B}^+\}$ and $D^- = \{(\Delta \mathbf{w}_r^B, \mathsf{False}) |  B \in \mathbb{B}^-\}$,  where $\Delta \mathbf{w}_r^B$ is the update of model $T_{r-1}$ computed on batch $B$. 
For misbehaving detection, $D^- = \{(\Delta \mathbf{w}_r^B, \mathsf{False}) |  B \in \mathbb{B}^-\}$ is the set of faithfully computed updates, and $D^+ = \{(\Delta \mathbf{\overline{w}}_r^B, \mathsf{True}) |  B \in \mathbb{B}^+\}$ where $\Delta \mathbf{\overline{w}}_r^B$ is defined according to the actual poisoning attack to be detected: $\Delta \mathbf{\overline{w}}_r^B = -\Delta \mathbf{w}_r^B$ for a gradient inversion attack, whereas $\Delta \mathbf{\overline{w}}_r^B$ is obtained by maximizing the loss function on  $B$ for a gradient ascent attack (see Section \ref{sec:threat_model}). 

\subsubsection{Property inference:} The detector model $M_r$ is applied on the reconstructed expected update $\Delta \mathbf{\widetilde{w}}_i$ for every client $i$, which results in $r$ individual decisions per client. 
These decisions are averaged to obtain the final decision about the property of each client.


\subsection{\our: Property reconstruction from linear features }
\label{sec:proposal}

The above technique applies OLS to reconstruct every single  coordinate of the expected update vector separately. Since the detector model $M_r$ combines every reconstructed gradient coordinate into a single decision, the reconstruction error per coordinate can accumulate and impact the decision, especially if $z$ is large.

We instead propose to first reconstruct $t$ linear features of every individual update vector ($t \ll z$), that capture the relevant property information, and then to infer the property values in this linear feature space. As aggregation is also a linear operation, gradient disaggregation corresponds to feature disaggregation in the feature space, therefore property inference can also be executed in this linear subspace of the gradient vectors  with an error that is proportional to $t$ (instead of $z$).






To make it more concrete, let $\tau_i \in \{0,1\}$ denote a binary variable indicating whether client $i$ has property $P$. 
Our goal is to find the property assignment $\bm{\tau}_{\max} = (\tau_1, \ldots \tau_N)$ with the largest likelihood $L(\bm{\tau} | \mathbf{b}_1, \ldots, \mathbf{b}_n)$ 
given the observed gradient aggregates as constraints, that is, $\bm{\tau}_{\max} = \arg\max_{\bm{\tau}} L(\bm{\tau} | \mathbf{b}_1, \ldots, \mathbf{b}_n)$. Let $g_r : \mathbb{R}^z \rightarrow \mathbb{R}^t$ be a linear function
that maps the update vector from the larger gradient space into a smaller feature space
where the property inference of an update is still accurate. In other words, $g_r$ performs feature reduction so that property-relevant information is preserved. In that case, the above likelihood maximization in the gradient space (given the gradient aggregates)  is roughly equivalent to likelihood maximization in the feature space (given the feature aggregates) due to the linearity of aggregation:
\begin{align}
\label{eq:final}
&\bm{\tau}_{\max} = \argmax_{\bm{\tau}} L\left(\bm{\tau} \vert \mathbf{b}_1, \ldots, \mathbf{b}_n\right) \notag \\
&\approx \argmax_{\bm{\tau}, \mathbf{X} \in \mathbb{X}} \prod_{i} \Bigg( \tau_i \cdot \prod_r p(\mathbf{X}_{r,i} | \tau_i = 1) 
   \notag \\ 
    & \hspace{1cm} + (1-\tau_i)\cdot \prod_r p(\mathbf{X}_{r,i} |  \tau_i = 0)\Bigg) 
\end{align}
where $\mathbf{X}_{r,i} \approx g_r(\Delta \mathbf{w}_r^i)$ is the individual feature vector of client $i$ in round $r$ whose per-round aggregates are given as constraints: $\mathbb{X} = \{\mathbf{X} \in \mathbb{R}^{n \times N \times t}\,|\,  \sum_{i} \mathbf{A}_{r,i} \mathbf{X}_{r,i} = g_r(\mathbf{b}_r)\}$, and $p(\mathbf{X}_{r,i}|\cdot)$ is approximated on the auxiliary data $D^{aux}$.
Owing to the linearity of $g_r$, the server can easily compute the feature aggregates by applying $g_r$ on the observed gradient aggregates $\mathbf{b}_r$:
\begin{align}
\label{eq:linearity}
g_r(\mb_r) = g_r\left(\sum_i^N \mathbf{A}_{r,i}\Delta\mathbf{w}_r^i\right)  = \sum_i^N \mathbf{A}_{r,i} \cdot g_r(\Delta\mathbf{w}_r^i) \notag \\ 
\approx \sum_i^N \mathbf{A}_{r,i} \cdot \mathbf{X}_{r,i}
\end{align}
Therefore, the server can  solve Eq.~\eqref{eq:final} and find a slightly biased approximation of $\bm{\tau}_{\max}$ in the feature space jointly with the most likely disaggregation $\mathbf{X}$ of the known feature aggregates (see Appendix \ref{sec:analysis} for a more detailed argument).

However, the individual features $\mathbf{X}$ are unobserved, other than their per-round aggregates, therefore the
above likelihood maximization is overly complex: Eq.~\eqref{eq:final} has a large number of variables ($\mathbf{X}$ and $\bm{\tau}$) and much fewer observations ($g_1(\mathbf{b}_1), \ldots, g_n(\mathbf{b}_n)$). This would yield an inaccurate approximation of $\bm{\tau}_{\max}$ even if $t$ is small.
Hence we introduce additional  constraints for the purpose of regularization:
The server computes the expected feature vector of a client from the known feature aggregates with linear regression and requires that these expected feature vectors match the mean $\sum_r \mathbf{X}_{r,i}/n$ of the reconstructed individual feature vectors of the same client. 
Linear regression is less likely to overfit with $t\cdot N$ variables, especially if $N < n$,  therefore can provide realistic constraints for the optimization problem in Eq.~\eqref{eq:final} and decrease the variance of its solution (see Appendix \ref{sec:comparison}).

Although the approximation of $\bm{\tau}_{\max}$  is biased in the feature space, it has a smaller variance than in the gradient space, which can eventually outbalance the bias and result in a more accurate property inference. This is detailed in Appendix \ref{sec:analysis} and also shown empirically in Section \ref{sec:evaluation}.
Indeed, Eq.~\eqref{eq:final} has fewer variables in the feature space 
and the regression can also be more accurate in this $t$-dimensional space.
We stress that \emph{the accurate approximation of $\bm{\tau}_{\max}$ in the feature space is only feasible because $g_r$, as well as gradient aggregation, are linear}. 

Our proposal has four main steps, which are also summarized in Table \ref{tab:prolin}: 

\begin{enumerate}
\item \textbf{Training the linear feature extractor $g_r$:}
The server learns the per-round feature extractor $g_r$  on $D^{aux}$.

\item \textbf{Computing the distribution of linear features:} 
The conditional probabilities $p(\mathbf{X}_{r,i}|\tau_i)$ in Eq.~\eqref{eq:final} are approximated with 
the client-independent feature distribution in round $r$, that is, the output distribution of $g_r$  on the held-out data $D^{aux}$.

\item \textbf{Reconstructing the expected linear features:}  For the purpose of regularization, the expected linear feature vector of every client is reconstructed from the feature aggregates $g_1(\mathbf{b}_1), \ldots, g_n(\mathbf{b}_n)$ with linear regression.


\item \textbf{Property inference:} Given the feature distributions from Step 2 and the reconstructed expected features per client from Step 3, the most likely property assignment $\bm{\tau}_{\max} = \arg\max_{\bm{\tau}} L(\bm{\tau} | \mathbf{b}_1, \ldots, \mathbf{b}_n)$ is approximated by solving  Eq.~\eqref{eq:final}.

\end{enumerate}

\subsubsection{Training the feature extractor}
\label{sec:training}


The server trains a detector model $M_r = h_r \circ g_r$ per round, which first extracts $t$ linear features of the update by applying $g_r: \mathbb{R}^z \rightarrow \mathbb{R}^t$ on the update vector and then applies a non-linear function $h_r : \mathbb{R}^t \rightarrow \{0,1\}$ on these linear features to recognize property $P$. Since the output of $h_r$ is the tested property, $h_r$ pushes $g_r$ to capture the property relevant information from the update vector. 
For example, if $g_r$ is a scalar linear function and $h_r$ is the sigmoid function, then $M_r(\mathbf{x}) = 1 / (1 + \exp(-g_r(\mathbf{x})))$ defines logistic regression. In that case, only a single feature is extracted from the entire update vector ($t=1$).

To train $M_r$, the server creates training data $D' = D^+ \cup D^-$, which consists of model updates with ($D^+$) and without ($D^-$) property $P$ just as described in Section \ref{sec:train_Mr_naive}. 
After splitting $D'$ into a training and testing part, the server trains $M_r$ on the training part of $D'$.




\begin{figure*}
    \centering
    \begin{subfigure}[]{0.238\textwidth}
        \includegraphics[width=\textwidth]{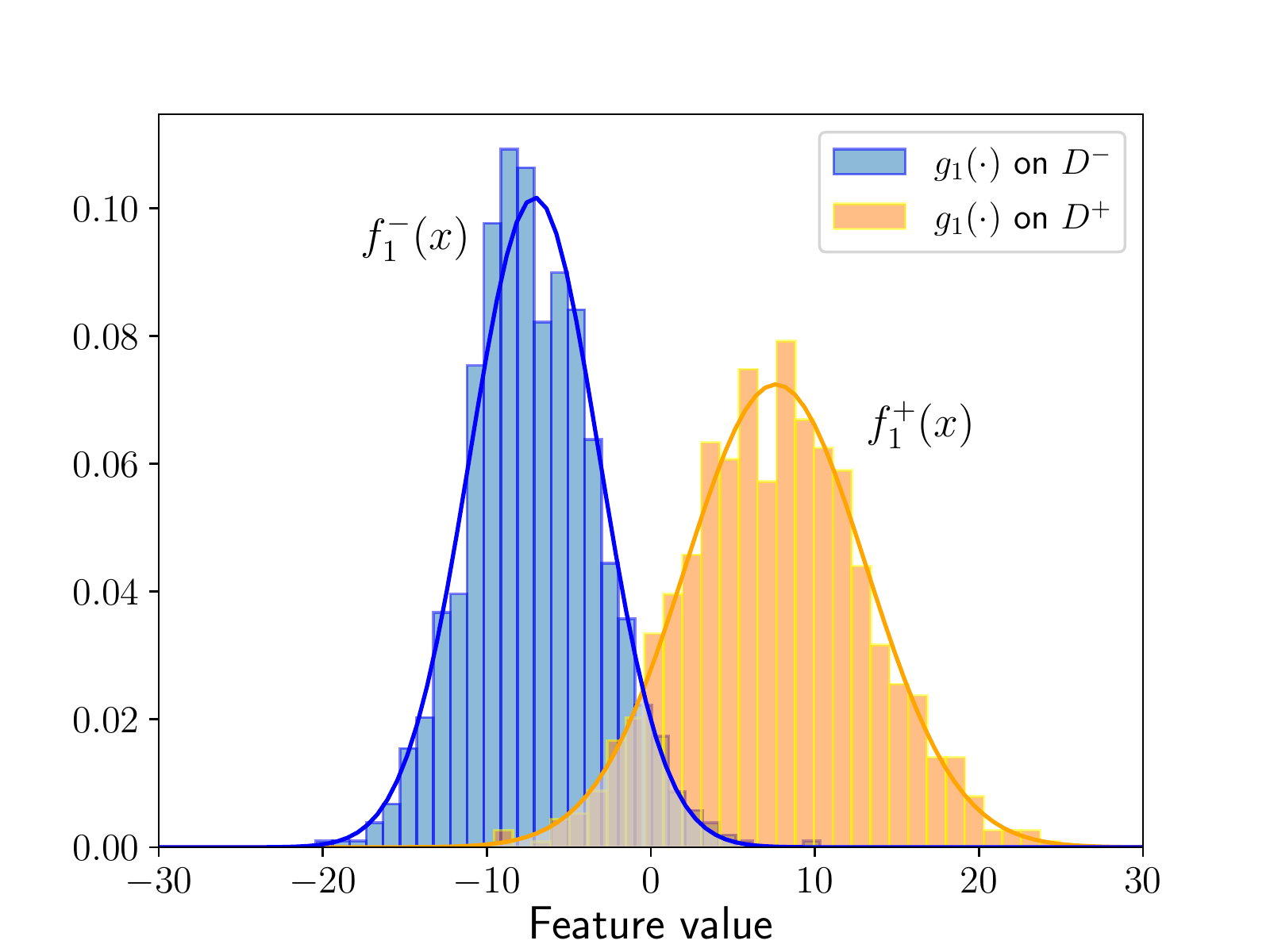}
        \caption{$r=1$}
    \end{subfigure}
    \begin{subfigure}[]{0.238\textwidth}
        \includegraphics[width=\textwidth]{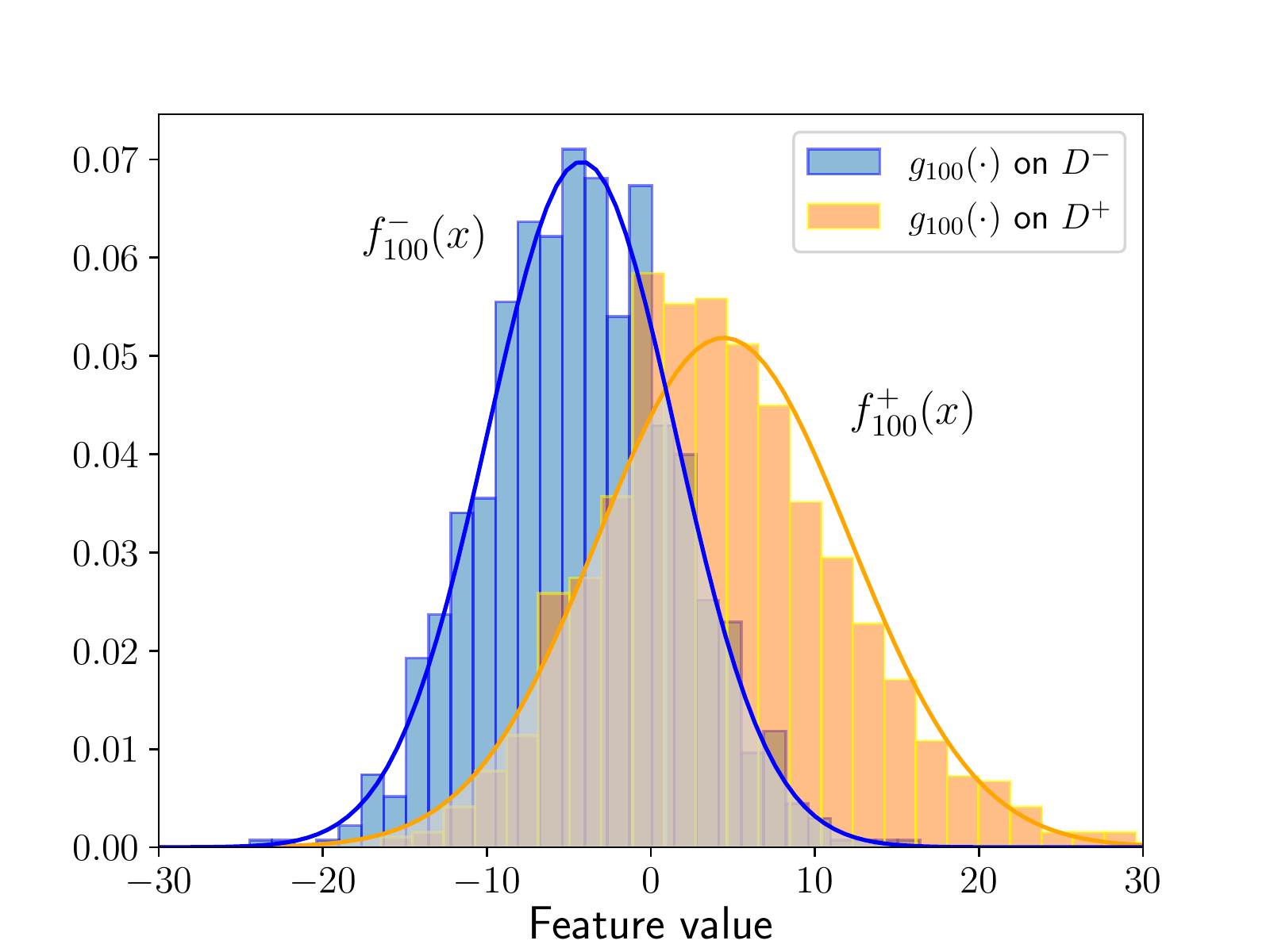}
        \caption{$r=100$}
    \end{subfigure}
    \vspace{-0.08cm}
    \begin{subfigure}[]{0.238\textwidth}
        \includegraphics[width=\textwidth]{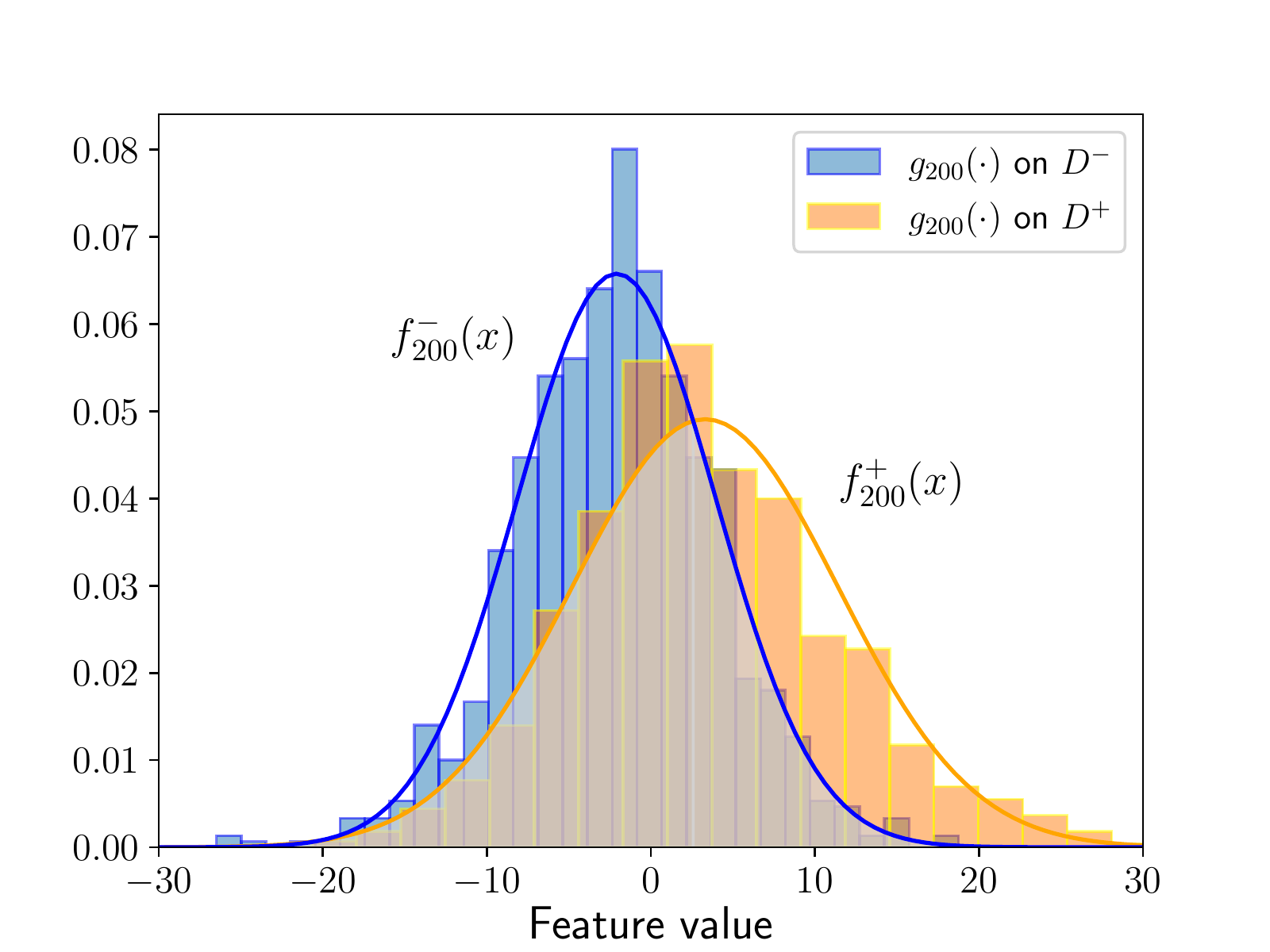}
        \caption{$r=200$}
    \end{subfigure}
    \vspace{-0.08cm}
    \begin{subfigure}[]{0.238\textwidth}
        \includegraphics[width=\textwidth]{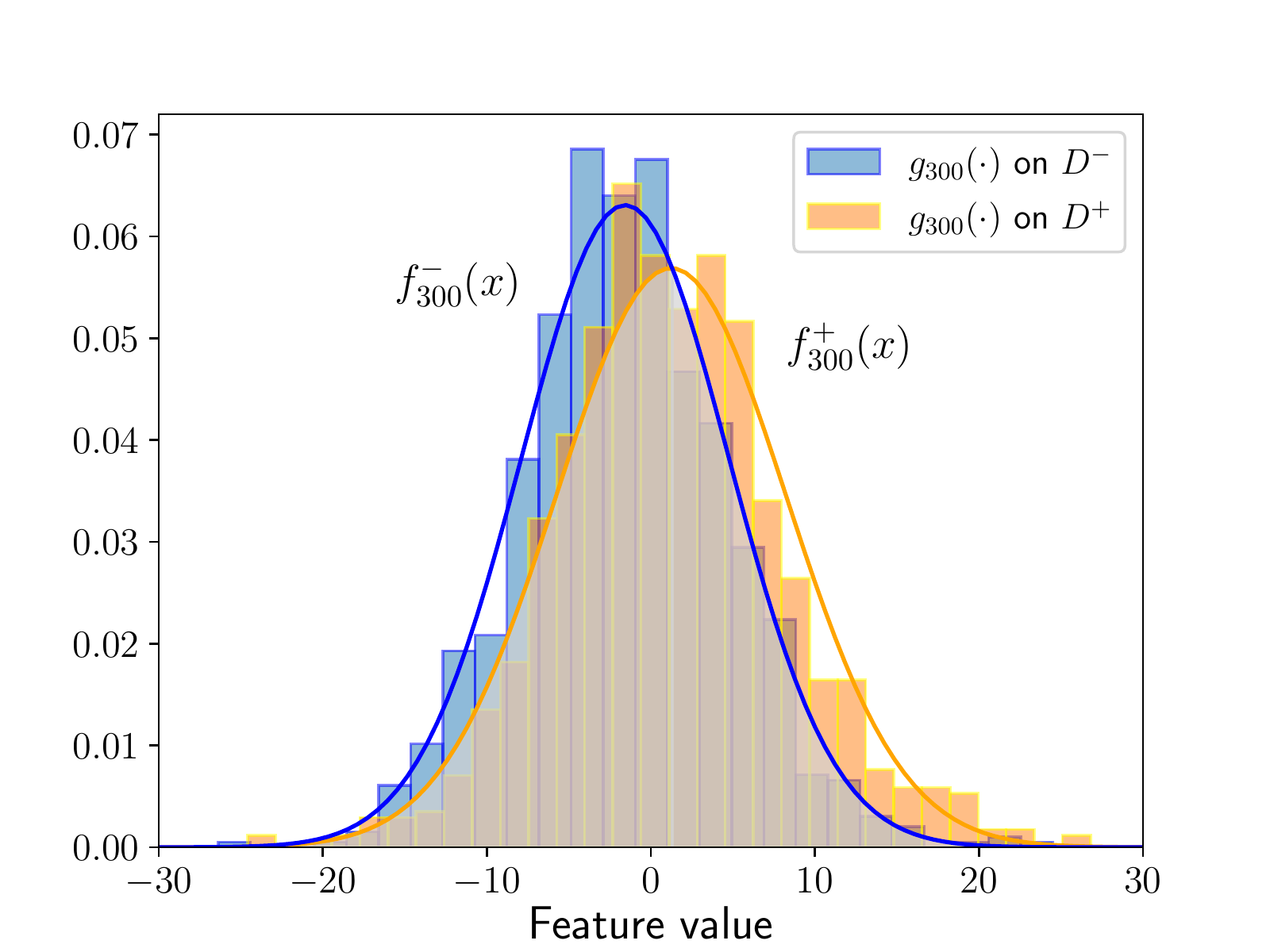}
        \caption{$r=300$}
    \end{subfigure}
    \caption{Distribution of a single linear feature $g_r(\Delta \mathbf{w}_r^B)$  for $B \in D^{aux}$ for membership detection depending on the round $r$ on MNIST ($t=1$). Overlap coefficient (OVL) is the overlap area between $f^+_r$ and $f^-_r$ colored with gray, measuring the performance of detector model $M_r$. }
    \label{fig:feature_dist}
\end{figure*}

\subsubsection{Computing the distribution of linear features}
\label{sec:feature_dist}
The output distributions of $g_r$ conditioned on $P$  are approximated on the testing part of $D'$ in every round $r$: 
$f^+_r$ denotes the Probability Density Function (PDF) of a random variable describing the output of $g_r$ on $D^+$, and $f_r^-$ denotes the PDF of a random variable describing the output of $g_r$ on $D^-$. As the server has no client-specific background knowledge to compute $p(\mathbf{X}_{r,i} | \tau_i)$  in Eq.~\eqref{eq:final}, it approximates $p(\mathbf{X}_{r,i} | \tau_i=1)$ with $f^+(\mathbf{X}_{r,i})$ and $p(\mathbf{X}_{r,i} | \tau_i=0)$ with $f^-(\mathbf{X}_{r,i})$. 

The distributions of a single linear feature are illustrated in Figure \ref{fig:feature_dist} for membership detection, where both $f_r^+$ and $f_r^-$ have normal distributions.
Figure \ref{fig:feature_dist} shows that $f_r^+$ and $f_r^-$ are well-separated at the beginning of the training, indicating an accurate detector model $M_r$. However, as the training progresses, the two distributions start to overlap, which implies a less accurate prediction.  In fact, for membership inference, $M_r$ is usually more accurate at the beginning of the training when the gradients are larger and the target sample  is likely to have a noticeable impact on the gradient. 
Unlike the naive approach in Section \ref{sec:baseline} that averages the output of $M_r$ over the rounds, \our  considers these potentially different per-round distributions of the features for more accurate inference. 

If the output of $g_r$ is high-dimensional (e.g., $t>1$) then the above approximation of $g_r$ with sampling becomes inaccurate unless $D'$ is sufficiently large. In that case, estimating $p(\mathbf{X}_{r,i} | \tau_i)$ with the output distribution of the entire detector model $M_r = h_r \circ g_r$  can be a more accurate approach. Since $h_r$ is designed to reduce the dimensionality of $g_r$, its output distribution conditioned on $P$ should provide a fairly accurate approximation of the feature distributions $f^+_r$ and $f^-_r$. 

\subsubsection{Reconstructing the expected linear features}
\label{sec:recon_features}
As the server can compute the feature aggregates based on Eq.~\eqref{eq:linearity}, it can also disaggregate them into the expected linear feature vector per client, similarly to gradient disaggregation in Section \ref{sec:gradient_recon}. 
However, instead of ordinary least square, we use a weighted version of ridge regression to address the potentially large variance of linear features.

More precisely, suppose that the aggregation of linear features can be described as 
\begin{align}
g_r(\mb_r) = \sum_i^N \mathbf{A}_{r,i} \cdot g_r(\Delta\mathbf{w}_r^i) = \sum_i^N \mathbf{A}_{r,i} \cdot \hat{\mathbf{g}}_i + \bm{\theta}_r
\label{eq:lin_model}
\end{align}
where $\hat{\mathbf{g}}_i \in \mathbb{R}^{t}$ is the vector of expected linear features 
of client $i$ that we want to reconstruct, and $\bm{\theta}_r \in \mathbb{R}^t$ is a vector of random noise that accounts for the stochasticity of learning and models the variance of the individual feature vectors over the rounds.  
Since the variance of $\bm{\theta}$ can be large, or the number of observations may be less than the number of features to recover ($n < t$), we use ridge regression with $L_2$ regularization of the reconstructed linear features  (see Section \ref{sec:lin_reg}): 
\begin{align}
\label{eq:prop_inf_grad}
\mathbf{\widetilde{G}} = \argmin_{\mx \in \mathbb{R}^{N \times t}} \mathbf{v} || \mathbf{G} - \mA \mx||_F^2 + \mathbf{\lambda} ||\mathbf{x}||_F^2
\end{align}
where $\mathbf{\widetilde{G}} \in \mathbb{R}^{n \times t}$ is an approximation of $\mathbf{\hat{G}} = (\mathbf{\hat{g}}_1, \ldots, \mathbf{\hat{g}}_N) \in \mathbb{R}^{n \times t}$, $\mathbf{G} = (g_1(\mb_1), \ldots, g_n(\mb_n)) \in \mathbb{R}^{N \times t}$, and $\mathbf{v}_r \in [0,1]$ ($1 \leq r \leq n$) denotes some measure of the performance of $M_r$. If $M_r$ provides an accurate prediction of property $P$, then the residual error $||g_r(\mb_r) - \sum_i \mathbf{A}_{r,i} \mathbf{x}_i||_2$ in round $r$ should have larger weight in the objective function because the output of $g_r$ is likely to have smaller variance.
Eq.~\eqref{eq:prop_inf_grad} can be solved efficiently by solving the objective function in Eq.~\eqref{eq:ridge} for each linear feature individually.


\subsubsection{Property inference}
\label{sec:prop_inference_prolin}
Given the feature aggregates $g_1(\mathbf{b}_1), \ldots, g_n(\mathbf{b}_n)$ from Eq.~\eqref{eq:linearity} and the reconstructed expected feature vectors  $\widetilde{\mathbf{G}}$ from Eq.~\eqref{eq:prop_inf_grad}, the server solves the following regularized version of  Eq.~\eqref{eq:final}: 
\begin{align}
\max_{\tau_i, \mathbf{X}} \quad &  \sum_{i} \log\left( \tau_i \cdot \prod_r f_r^+(\mathbf{X}_{r,i}) 
    + (1-\tau_i)\cdot \prod_r f_r^-(\mathbf{X}_{r,i})\right) \notag \\
\textrm{s.t.} \quad & \: \sum_{i} \mathbf{A}_{r,i} \mathbf{X}_{r,i} = g_r(\mathbf{b}_r)  \tag{Constraint 1} \\
 & \: \sum_{r \in R(i)} \frac{\mathbf{X}_{r,i}}{|R(i)|}  = \widetilde{\mathbf{G}}_i  \tag{Constraint 2} \\
  & \: \tau_i \in \{0,1\}, \mathbf{X} \in \mathbb{R}^{n \times N \times t}   \tag{Constraint 3}
\end{align}
where $f_r^+$ and $f_r^-$ denote the PDFs of the linear features conditioned on property $P$  (see Section \ref{sec:feature_dist}), and $R(i)$ is the set of rounds in which client $i$ participates.
Constraint 1 requires that the reconstructed linear features $\mathbf{X}_{r,i}$ should produce the known feature aggregates. 
Constraint 2 provides regularization by requiring that the mean of the reconstructed linear features $\mathbf{X}_{r,i}$ should match the expected linear feature vector per client  (see Eq.~\eqref{eq:lin_model} and Eq.~\eqref{eq:prop_inf_grad}). Finally, Constraint 3 pushes the optimization to find an integer-valued solution for $\tau_i$, as a client either has or does not have property $P$. 

The above optimization problem contains integer variables  $\tau_i$, which makes the problem NP-complete. Hence, we relax the problem into
\begin{small}
 \begin{align}
    \min_{\tau_i, \mathbf{X}} \quad &\gamma_1 \mathcal{L}_{\mathsf{ml}} + \gamma_2 \mathcal{L}_{\mathsf{reg}} + \gamma_3 \mathcal{L}_{\mathsf{lstsq}} \notag \\
    \textrm{where} \quad & \mathcal{L}_{\mathsf{ml}} = - \sum_{i} \log\left( \tau_i \cdot \prod_r f_r^+(\mathbf{X}_{r,i}) +  (1-\tau_i)\cdot \prod_r f_r^-(\mathbf{X}_{r,i})\right) \notag \\
    & \mathcal{L}_{\mathsf{reg}} = \sum_{i,r \in R(i)} \left\lVert\frac{\mathbf{X}_{r,i}}{|R(i)|}  - \widetilde{\mathbf{G}}_i\right\rVert^2_F \notag \\
    & \mathcal{L}_{\mathsf{lstsq}} =  \sum_{i,r} \mathbf{v}_r\lVert g_r(\mathbf{b}_r) - \mathbf{A}_{r,i} \mathbf{X}_{r,i}\rVert_F^2  \notag \\
    \textrm{s.t.} \quad &  0 \leq \tau_i \leq 1, \mathbf{X} \in \mathbb{R}^{n \times N \times t}   \notag
    \end{align}
\end{small}
where $\gamma_1$, $\gamma_2$, $\gamma_3$ are the weighting factors of each loss in the objective function. Although this relaxed version is still non-convex if $f_r^-$ or $f_r^+$ are also non-convex, the variable $\tau_i$ is now continuous in $[0,1]$ and hence  can be approximated with projected gradient descent (e.g., using an automatic differentiation framework such as PyTorch \cite{paszke2019pytorch}).

\begin{table}
\begin{tcolorbox}
\caption{\our}
\label{tab:prolin}
\textbf{Input:} 
 (1) observed aggregate $\mb_r = \sum_i^N \mathbf{A}_{r,i} \cdot \Delta\mathbf{w}_r^i$ per round; (2) participation matrix $\mA$; (3) auxiliary data $D^{\mathit{aux}}$; (4) property $P$; (5) federated model $T_r$ per round \smallskip \\
\textbf{Output:} $\bm{\tau} \in \{0,1\}^N$ ($\bm{\tau}_i = 1$ if client $i$ has property $P$) \medskip \\
\textbf{Begin}
\begin{enumerate}
    \item \textbf{Training property inference model:} For every round $r$ use auxiliary data $D^{aux}$ to
    \begin{enumerate}
    \item train detector model $M_r = h_r \circ g_r$  
    \item compute weight $\mathbf{v}_r$, which is proportional to some performance metric of $M_r$ 
    \end{enumerate}
    \item \textbf{Computing the distribution of linear features:} Approximate the PDFs $f^+_r$ and $f^-_r$  of the linear features $g_r$ conditioned on property $P$ on $D^{aux}$
    \item \textbf{Reconstructing expected linear features:} 
    \begin{enumerate}
    \item Compute the aggregation of linear features $\mathbf{G}_r$ in every round $r$: 
    \begin{align*}
    \mathbf{G}_r = g_r(\mb_r) = \sum_i^N \mathbf{A}_{r,i} \cdot g_r(\Delta\mathbf{w}_r^i)
    \end{align*}
    \item Approximate the expected linear features $\sum_{r \in R(i)}\frac{g_r(\Delta \mathbf{w}_r^i)}{|R(i)|}$ per client $i$ by $\mathbf{\widetilde{G}}_i$, where
    \begin{align*}
    \mathbf{\widetilde{G}} = \argmin_{\mx \in \mathbb{R}^{N\times t}} \mathbf{v} || \mathbf{G} - \mA \mx||_F^2 + \mathbf{\lambda} \cdot || \mathbf{x}||_F
    \end{align*}
    \end{enumerate}

     \item \textbf{Property inference:} Solve the following optimization problem for $\bm{\tau}$:
\begin{small}
 \begin{align}
    \min_{\tau_i, \mathbf{X}} \quad &\gamma_1 \mathcal{L}_{\mathsf{ml}} + \gamma_2 \mathcal{L}_{\mathsf{reg}} + \gamma_3 \mathcal{L}_{\mathsf{lstsq}} \notag \\
    \textrm{where} \quad & \mathcal{L}_{\mathsf{ml}} = - \sum_{i} \log \Bigg( \tau_i \prod_r f_r^+(\mathbf{X}_{r,i}) \notag \\ & \hspace{1cm}+ (1-\tau_i) \prod_r f^-_r(\mathbf{X}_{r,i})\Bigg) \notag \\
    & \mathcal{L}_{\mathsf{reg}} = \sum_{i,r \in R(i)} \left\lVert \frac{\mathbf{X}_{r,i}}{|R(i)|}  - \mathbf{\widetilde{G}}_i\right\lVert_F^2 \notag \\
    & \mathcal{L}_{\mathsf{lstsq}} =  \sum_{i,r} \mathbf{v}_r \left\lVert \mathbf{G}_r - \mathbf{A}_{r,i} \mathbf{X}_{r,i}\right\rVert_F^2  \notag \\
    \textrm{s.t.} \quad &  0 \leq \tau_i \leq 1, \mathbf{X} \in \mathbb{R}^{n \times N \times t}   \notag
    \end{align}
\end{small}
\end{enumerate}
\textbf{End}
\end{tcolorbox}
\end{table}


We provide a theoretical justification of \our in Appendix~\ref{sec:analysis}.

\section{Evaluation}
\label{sec:evaluation}
In this section, we demonstrate that \our can effectively disaggregate the linear features of different detector models. Although we focus on two specific detection tasks (membership inference and misbehaving detection), 
we emphasize that \our is a general approach that can disaggregate any linear function and hence potentially reconstruct various client-specific properties given their accurate detector models.  Moreover, even if the detector model is not consistently accurate in every round, \our takes the best combination of these per-round models to have a quasi-optimal property inference.  

\subsection{Dataset}

\label{sec:datasets}
We compare the performance of \our with different property reconstruction techniques. 
We evaluate all approaches on the following datasets:
\begin{itemize}
    \item The MNIST database of handwritten digits. It consists of 28 x 28 grayscale images of digit items and has 10 output classes. The training set contains 60,000 data samples, while the test/validation set has 10,000 samples \cite{MNIST}.
    \item The CIFAR-10 dataset consists of 60,000 32x32 color images in 10 classes, with 6000 images per class. There are 50,000 training images and 10,000 test images \cite{CIFAR}.
    \item Fashion-MNIST database of fashion articles consists of 60,000 28x28 grayscale images of 10 fashion categories, along with a test set of 10,000 images \cite{Fashion-MNIST}. 
\end{itemize}
For each dataset, we randomly select 10\% of the training set for auxiliary data $D^{aux}$. Therefore, the server has $|D^{aux}| = 6000$ samples for MNIST and Fashion-MNIST and $|D^{aux}| = 5000$  samples for CIFAR-10. $D'$ is generated from $D^{aux}$ as described in Section \ref{sec:training}, where $|D'| = 2\cdot |D^{aux}|$ in our evaluations\footnote{Since $D'$ contains batches of $D^{aux}$, it can have larger size.}. We use 80\% of $D'$ to train $M_r$ and 20\% to compute weights $\mathbf{v}$ in Eq.~\eqref{eq:prop_inf_grad} as well as distributions $f^+_r$ and $f^-_r$.

\subsection{Model Architectures}
As in \cite{lam2021gradient}, we use LeNet neural network as the global model $T_r$ with the following architectures:

\begin{itemize}
    \item For MNIST and Fashion-MNIST, we use two 5x5 convolution layers (the first with 10 filters, the second with 20), each followed by 2x2 max pooling, a dropout layer with ratio set to 0.5, and two fully connected layers with 50 and 10 neurons, respectively. A dropout layer separates the two fully connected layers.
    \item For CIFAR-10, we use two 5x5 convolution layers (the first with 6 filters, the second with 16), each followed by 2x2 max pooling and three fully connected layers with 120, 84 and 10 neurons, respectively.
\end{itemize}

\subsection{Property reconstruction}
We consider the detection of three properties: (1) membership information of a randomly chosen target sample, malicious behavior by launching (2) gradient inversion or (3) gradient ascent attack.

\subsubsection{Approaches}
We compare the following approaches to reconstruct the above properties.\medskip \\
\noindent \textbf{BASELINE}: This is based on gradient reconstruction from~\cite{lam2021gradient}, which is also described in Section~\ref{sec:baseline}. As opposed to~\cite{lam2021gradient}, we train a distinct inference model $M_r$ per round instead of a single model over all the rounds and average the decisions of these per-round models. This "ensemble" approach is more accurate since $M_r$ can have very different performances per round.   $M_r$ is a logistic regression model. \smallskip \\
\noindent \textbf{PROLIN}: 
This is  based on feature reconstruction and described in Section \ref{sec:proposal}. It is instantiated with a single linear feature ($t=1$) and $M_r = h \circ g_r$ is a logistic regression model where $h$ is the sigmoid function. As $M_r$ is trained only on the update of a single round, this simple model is accurate and also fast to train. 
Following from empirical observations (also illustrated in Figure \ref{fig:feature_dist}), the feature distributions $f^+_r$ and $f^-_r$ are approximated to be normal\footnote{The PDF of a normal random variable with mean $\mu$ and variance $\sigma^2$ is $\frac{1}{\sigma\sqrt{2\pi}}e^{\frac{-(x-\mu)^2}{2\sigma^2}}$} whose means and variances equal the empirical means and variances of the single linear feature $g_r$ on the testing part of $D'$ conditioned on property $P$.
The weight $\mathbf{v}_r$ per round is set to $1-\mathrm{OVL}_r$, where $\mathrm{OVL}_r$ is the overlapping coefficient between the distributions of $f_r^+$ and $f_r^+$
and is a value between 0 and 1 that measures the overlap area of the two probability density functions (also illustrated in Figure \ref{fig:feature_dist}). Therefore, a coefficient with a small value means that $M_r$ is accurate and vice versa\footnote{$1-\mathrm{OVL}$ is also equivalent to Youden's index since OVL is the sum of False Negative and False Positive Ratios}. 
The regularization parameter $\lambda$ is fixed to 5 for all experiments. The weights $\gamma_1$, $\gamma_2$, and $\gamma_3$ of different losses in the objective function of \our (see Section \ref{sec:prop_inference_prolin}) are adjusted dynamically during optimization using the technique in \cite{malkiel2020mtadam}.
\smallskip \\
\noindent \textbf{OLS}: This is based on feature reconstruction that infers the property by applying the sigmoid function $h$ 
on the approximation $\mathbf{\widetilde{G}}_i$ of the single expected linear feature of a client ($t=1$).  $\mathbf{\widetilde{G}}_i$ is obtained by solving Eq.~\eqref{eq:prop_inf_grad} with $\mathbf{v}=\mathbf{1}$ and $\lambda = 0$, that is,  each round has equal weight and there is no regularization. 
$\tau_i = 1$ if $h(\mathbf{\widetilde{G}}_i) > 0.5$. \smallskip \\
\noindent \textbf{REG}: This is based on a feature reconstruction like OLS, except that \emph{ridge regression} is applied to obtain $\mathbf{\widetilde{G}}_i$ by solving Eq.~\eqref{eq:prop_inf_grad} with $\mathbf{v}$ and $\lambda$ as defined in \our. 
    $\tau_i = 1$ if $h(\mathbf{\widetilde{G}}_i) > 0.5$. 

\subsubsection{Experimental setup}
We perform 3 runs for each experiment and average the results obtained over these 3 runs. For the global model, we use a batch size of 10 for MNIST and Fashion-MNIST and 25 for CIFAR-10 and a batch size of 10 to train $M_r$. The learning rate $\eta$ is set to 0.01 to train the global model for MNIST and Fashion-MNIST, 0.1 for CIFAR-10, and 0.001 to train $M_r$ in order to identify $g_r$. SGD is used to train all models.
For membership inference attack (MIA), the target sample is chosen uniformly at random in each experiment and remains fixed over all the rounds for one experiment. All clients with the membership property have this sample in their local training data. 
The number of federated rounds is $n=300$, the number of all clients is $N=50$, and a fraction of $\phi = 0.1$ of all clients are positive (i.e., have the property). In each round, a fraction of $C=0.2$ of all clients are selected uniformly at random to send their model update for aggregation after performing a single epoch of local training on their own training data.
Each client has the same number of training samples, which are assigned to the clients uniformly at random at the very beginning of the training.
All settings are summarized in Table \ref{tab:params} in the appendix. 


\subsection{Results}
\begin{figure*}[!ht]
\centering
\includegraphics[scale=0.35]{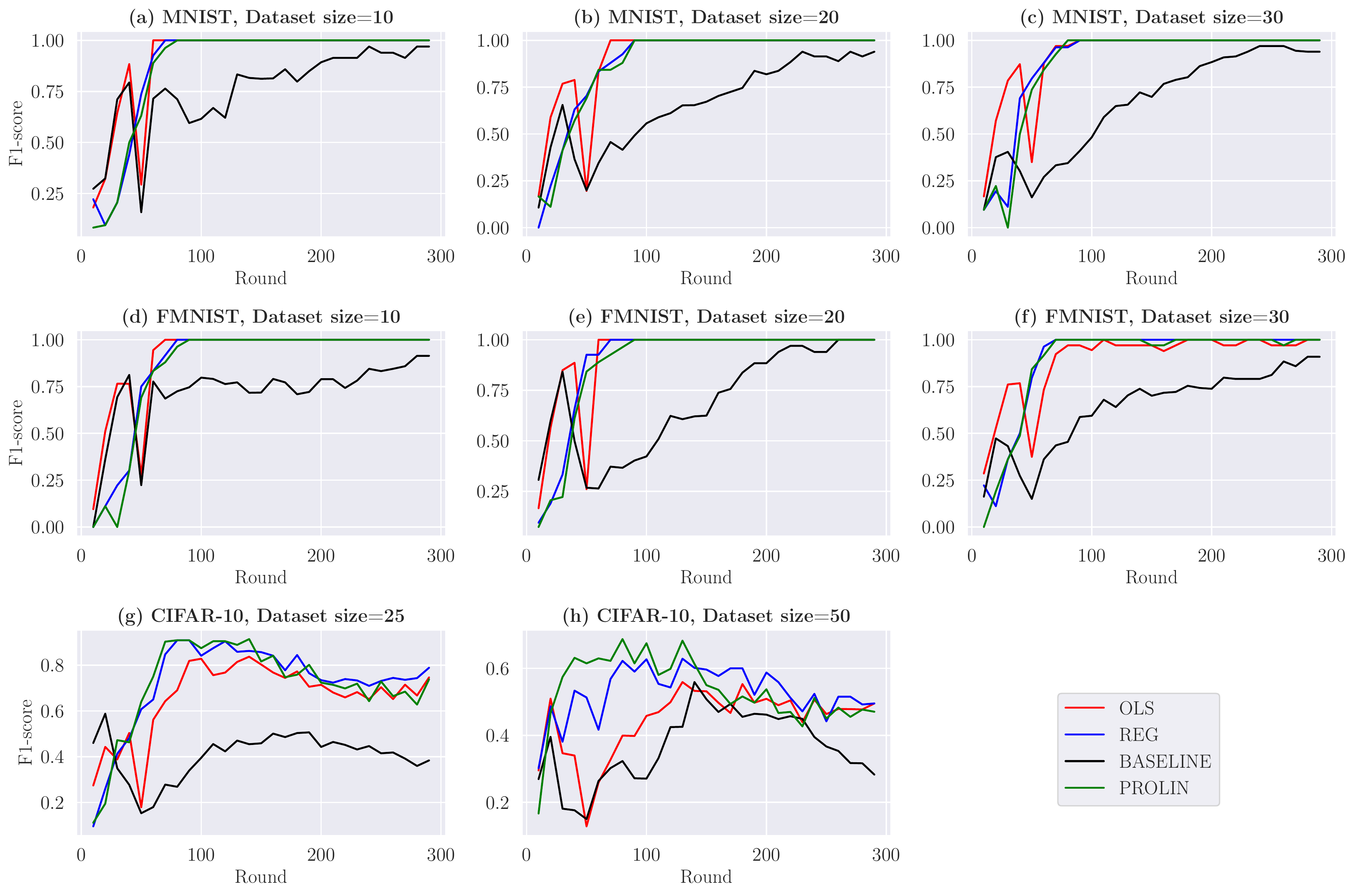}
\caption{Misbehaving detection (gradient inversion attack) on the MNIST, Fashion-MNIST, and CIFAR-10 datasets by varying the size of the local dataset per client ($|D_i|$)}
\label{fig:inv_attack}
\end{figure*}

\begin{figure*}[!ht]
\centering
\includegraphics[scale=0.35]{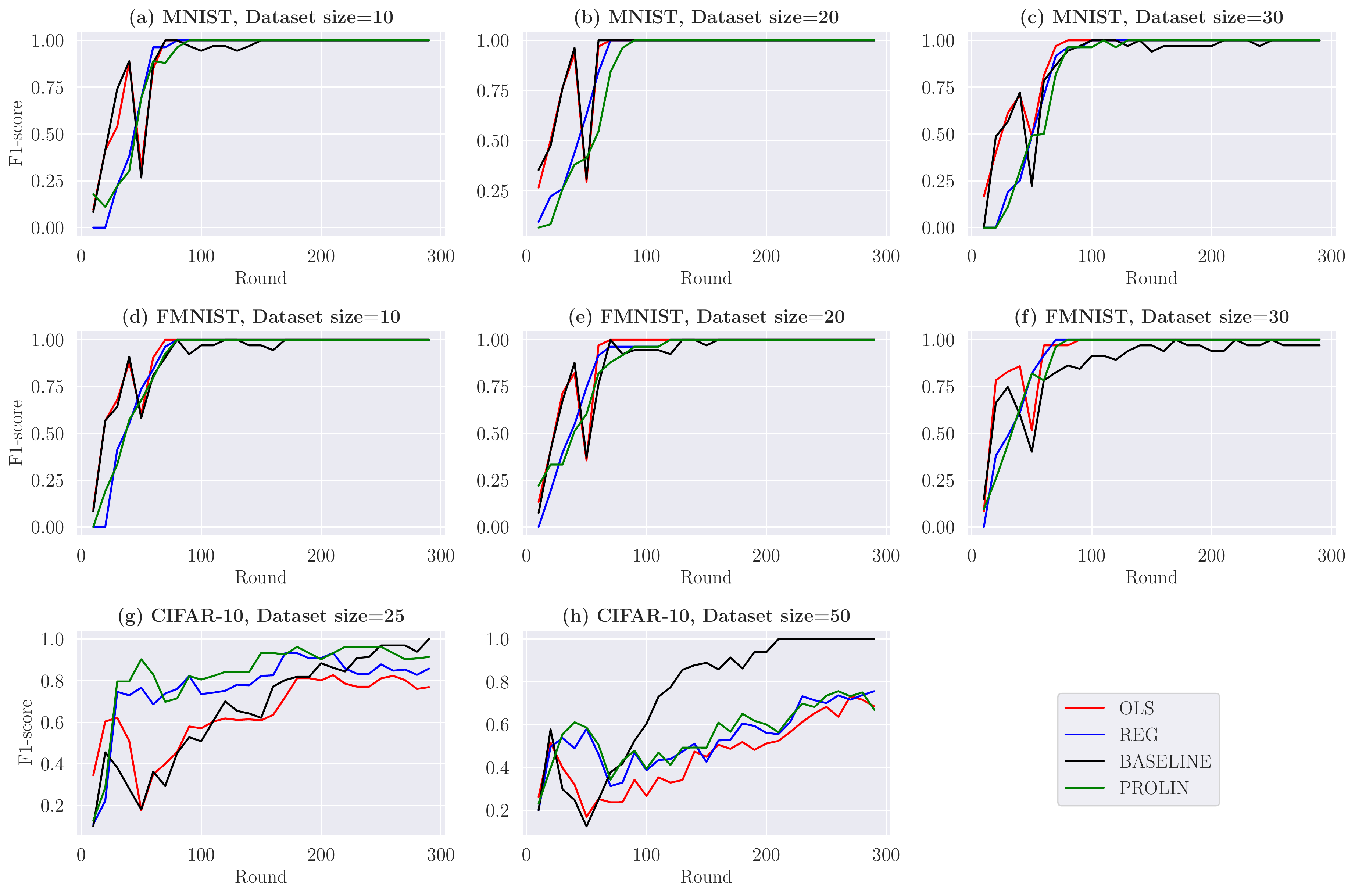}
\caption{Misbehaving detection (gradient ascent attack) on the MNIST, Fashion-MNIST, and CIFAR-10 datasets by varying the size of the local dataset per client ($|D_i|$)}
\label{fig:aga_attack}
\end{figure*}

Figures~\ref{fig:mia_attack},~\ref{fig:inv_attack} and ~\ref{fig:aga_attack} show the results for the membership inference attack (MIA) and for the detection of gradient inversion (INV) and gradient ascent attacks (GAA), respectively. 
We report the F1-score of the detection, which is the harmonic mean of the precision and recall, where the precision is the number of correctly detected positive clients divided by the number of all clients who are detected as positive, and the recall is the number of correctly detected positive clients  divided by the number of all positive clients.

Out of the three detection tasks, membership information is the most difficult (Figure~\ref{fig:mia_attack}) and gradient ascent is the easiest to detect (Figure~\ref{fig:aga_attack}).  Indeed, a single target sample has less significant impact on the aggregated model update as the training progresses (see Section \ref{sec:training}), while gradient manipulation depends only on the performance of the common model $T_{r-1}$. Unlike gradient ascent, gradient inversion does not modify the magnitude of the update, hence it is more difficult to detect (Figures~\ref{fig:inv_attack} and \ref{fig:aga_attack}).

\begin{figure*}[!ht]
\centering
\includegraphics[scale=0.35]{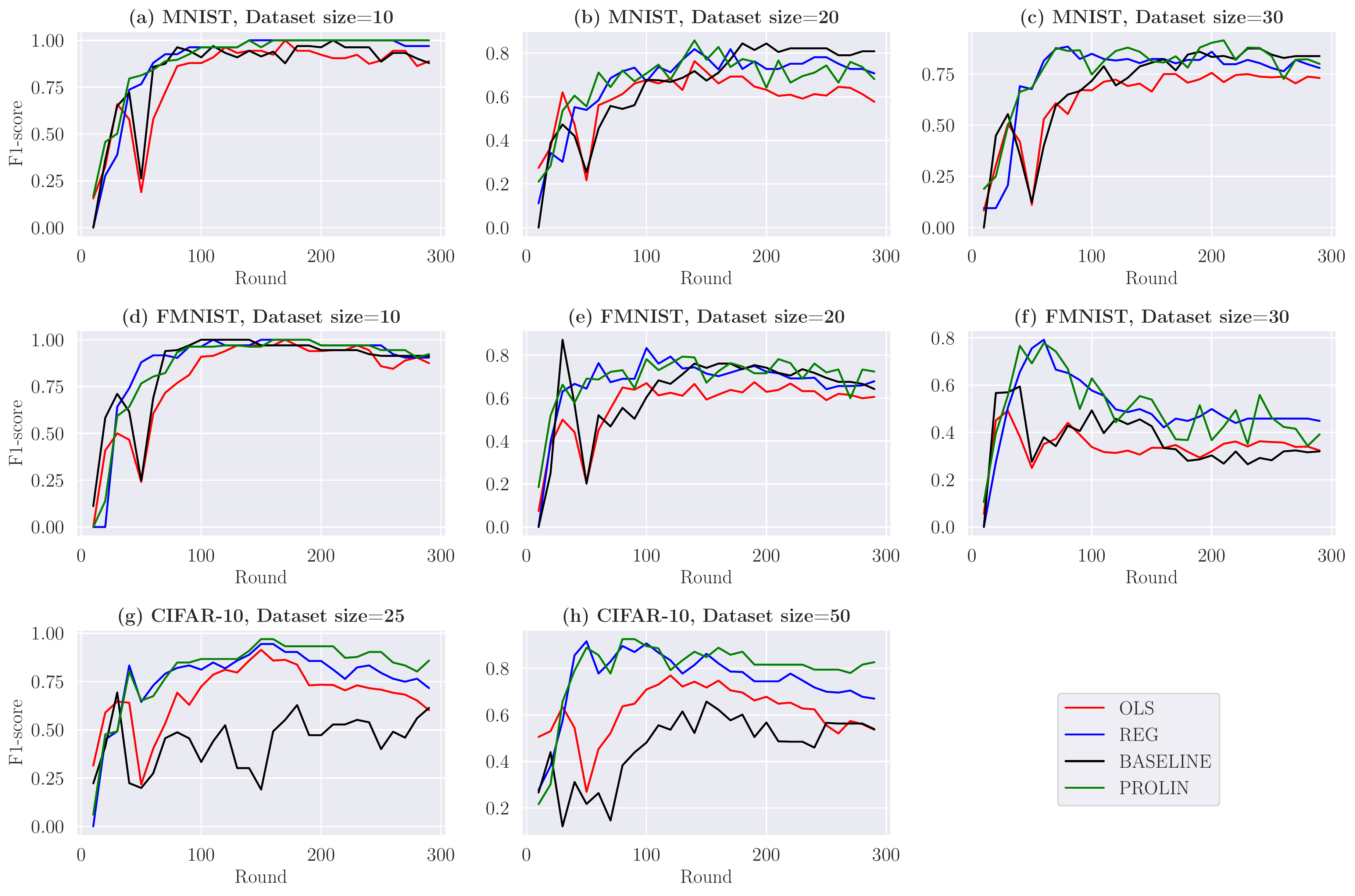}
\caption{Membership detection on the MNIST, Fashion-MNIST, and CIFAR-10 datasets by varying the size of the local dataset per client ($|D_i|$)}
\label{fig:mia_attack}
\end{figure*}

\subsubsection{Feature vs. gradient reconstruction}
\begin{figure}
    \centering
        \includegraphics[scale=0.26]{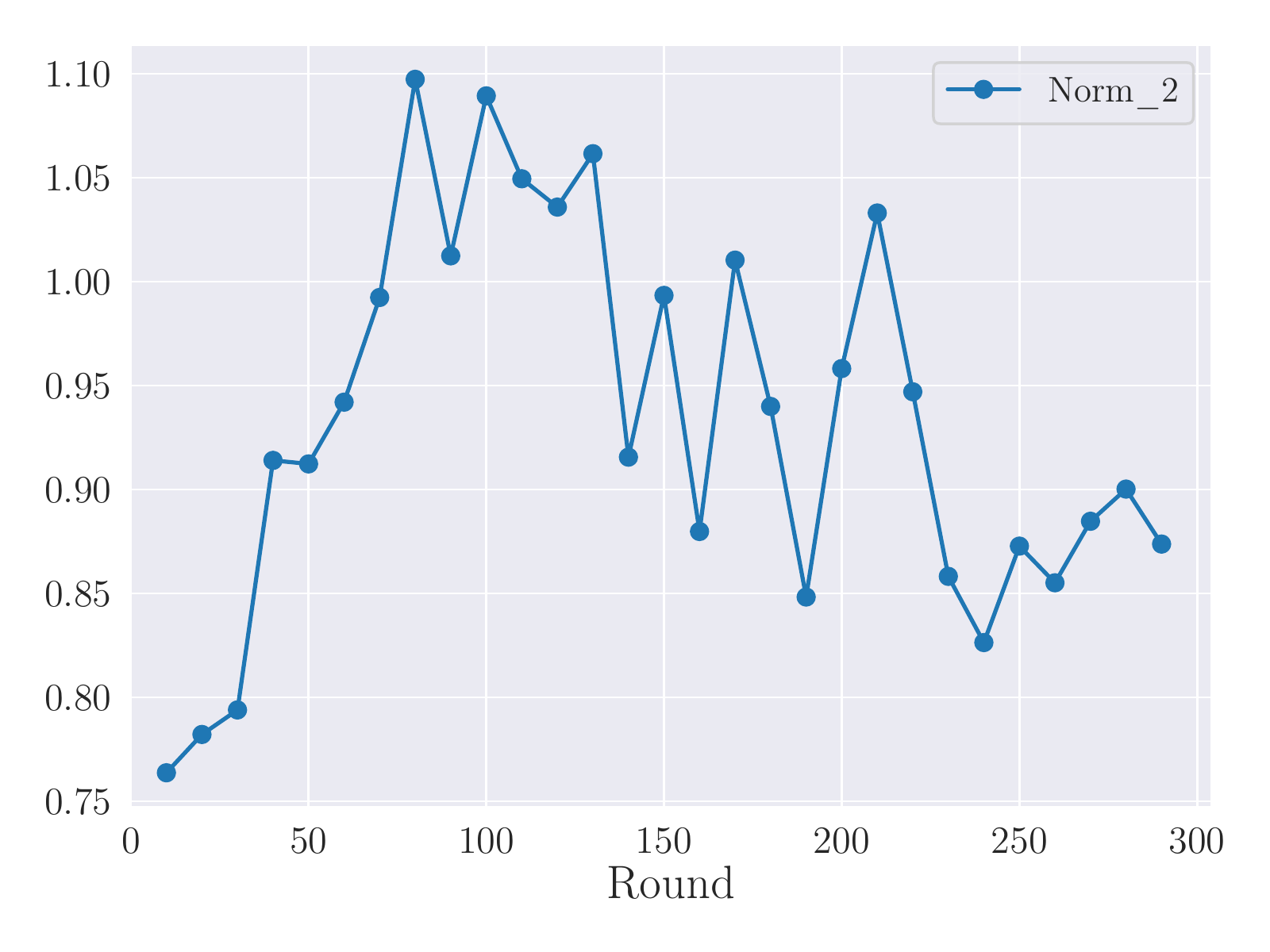}
        \caption{$L_2$-norm of the weights of the linear model $g_r$ on CIFAR10 with client datasize 50 when a gradient ascent attack is detected.}
    \label{fig:weight_norm}
\end{figure}

Feature reconstruction (OLS, REG, \our) is superior to gradient reconstruction (BASELINE) on almost all tasks, albeit to different degrees. 
The difference is the most salient on CIFAR-10, which shows that \textbf{feature reconstruction is indeed a more appealing approach for property inference especially if the common model is more complex}. The exception is the detection of the gradient ascent attack when the dataset size is $50$ (see Figure~\ref{fig:aga_attack}.h), where BASELINE is more accurate than other approaches. Indeed, Figure \ref{fig:weight_norm} depicts the $L_2$-norm of the weights $\bm{\alpha}$ of the linear model $g_r$ depending on $r$ for this scenario. This shows that the norm falls below 1 after round 100, which means that the reconstruction error of BASELINE can be less than for other approaches as explained in Appendix \ref{sec:comparison}. 
Feature reconstruction also generally shows a smaller variance in accuracy over the rounds, and it converges faster especially when  clients have larger datasets. 

\subsubsection{PROLIN vs. BASELINE}
For MIA, BASELINE reaches the maximum F1-score of 0.83 in Figure~\ref{fig:mia_attack}.e at round 20 and then drops quickly, while \our reaches the peak at round 120 with an F1-score of 79\%, providing a more stable performance. Similarly, in Figure~\ref{fig:mia_attack}.b, BASELINE is more accurate at the end of the training, however, \our obtained the maximum F1-score (86\%) in this case. In fact, BASELINE reaches the best performance with the GAA shown in Figure~\ref{fig:aga_attack}.h by reaching an F1-score of 100\%, while \our reaches 76\%. Nevertheless,  \our is more accurate overall and has larger F1-scores on average over the rounds. For example, the worst F1-score on all the considered scenarios is 69\% for \our, while it is 56\%  for BASELINE (Figure~\ref{fig:inv_attack}.h).
In addition, \textbf{\our has a more stable performance with a smaller variance over the rounds than BASELINE}. Indeed, in many cases, the F1-score of BASELINE has a larger variance (Figure~\ref{fig:mia_attack}.f-h, Figure~\ref{fig:inv_attack}.g-h), which makes it difficult to choose the round number where it provides good performance: Even if the detector can access all the rounds, it must choose one where it is supposed to obtain the final detection result. 
It is therefore crucial to have a stable performance to ensure that the reconstruction remains accurate over a sufficiently wide range of rounds. Finally, \textbf{\our also converges much faster to good F1-scores in general} (Figure~\ref{fig:mia_attack}.e-h, Figure~\ref{fig:inv_attack}.a-h, Figure~\ref{fig:aga_attack}.f-g). This is also important because the server stops federated learning as soon as the global model reaches acceptable performance, and therefore the reconstruction must also be accurate by then. 

\subsubsection{PROLIN vs. REG and OLS}
\begin{figure}[!ht]
    \centering
    \begin{subfigure}[]{0.22\textwidth}
        \includegraphics[width=\textwidth]{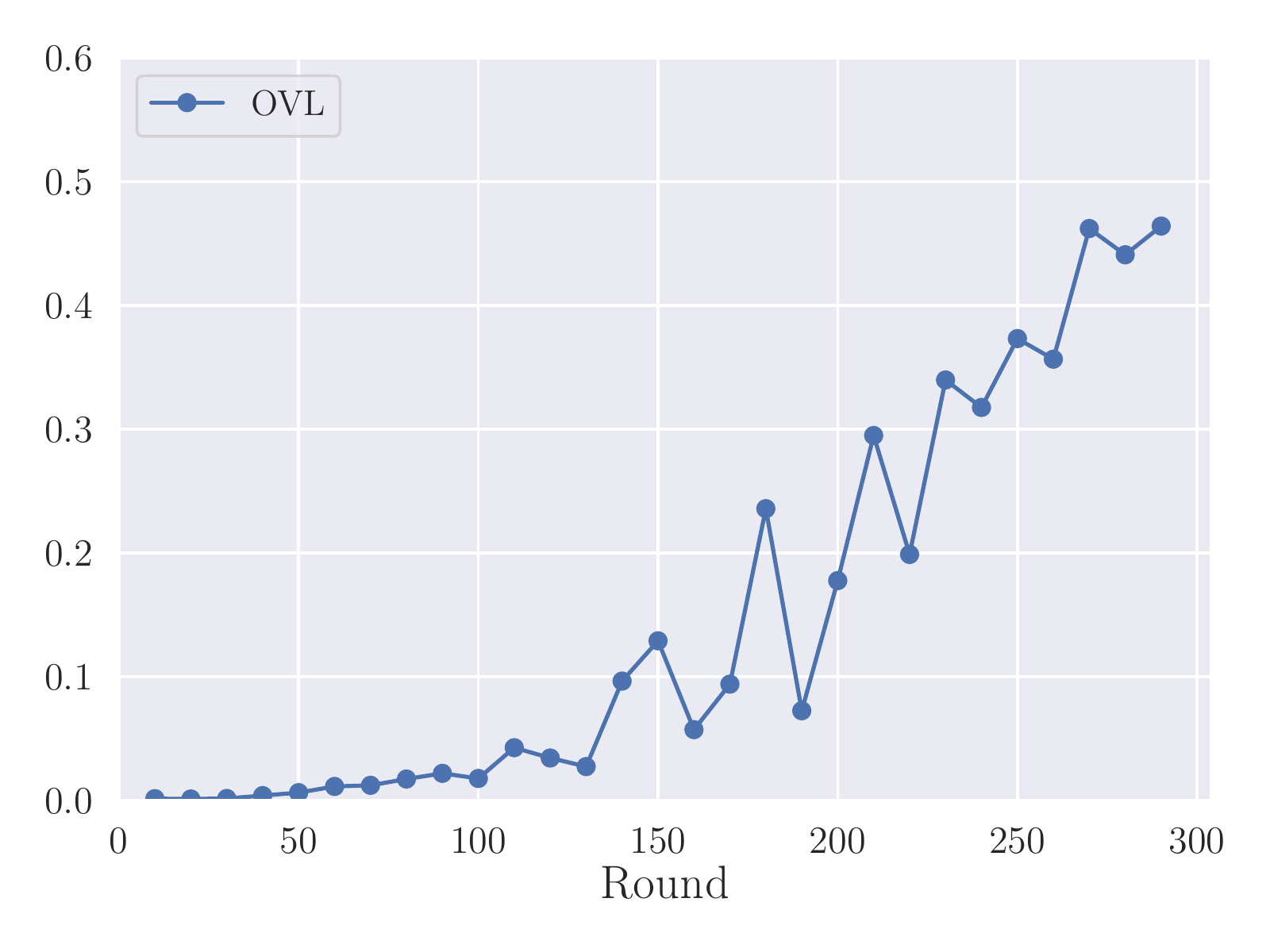}
        \caption{Membership inference attack}
    \end{subfigure}
    \begin{subfigure}[]{0.22\textwidth}
        \includegraphics[width=\textwidth]{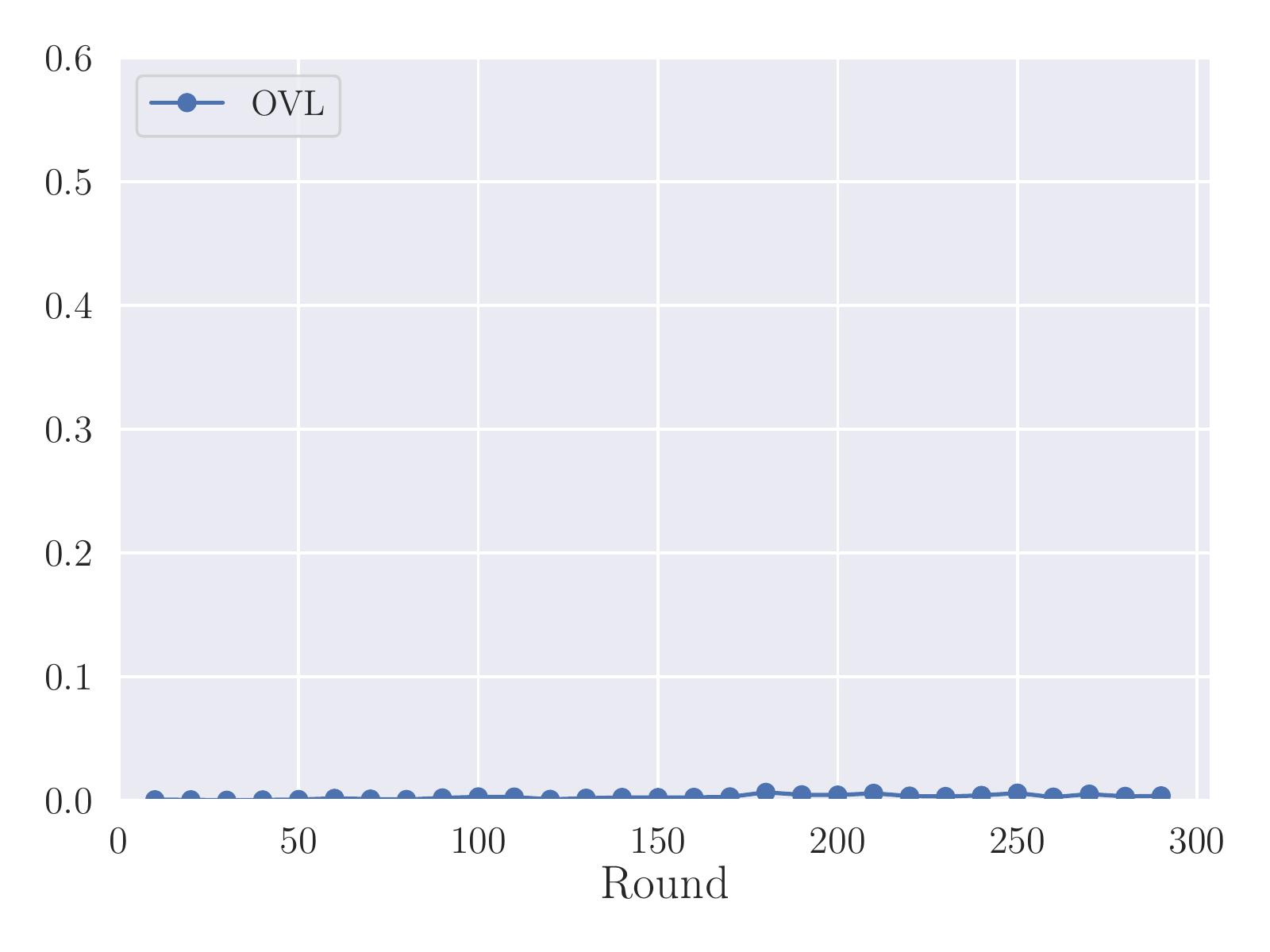}
        \caption{Gradient inversion attack}
    \end{subfigure}
    \vspace{-0.08cm}
    \caption{OVL coefficients over the rounds for membership and misbehaving detection on CIFAR-10 with dataset size 25.}
    \label{fig:OVL}
\end{figure}

\our is also superior to REG and OLS. Although this difference is more apparent when \our is compared with OLS (Figure~\ref{fig:mia_attack}.a-h, Figure~\ref{fig:inv_attack}.f-h, Figure~\ref{fig:aga_attack}.g-h), \our is also more accurate than REG over all the rounds when a MIA is executed on CIFAR-10  (Figure~\ref{fig:mia_attack}.g-h). For the other cases, they have very similar performance: 
 $\mathcal{L}_{\mathsf{ml}}$ evaluates the likelihood with the round specific feature distributions $f^+_r$ and $f^-_r$, while REG uses the fixed sigmoid function $h$ in all the rounds. Since $h$ infers the property from a single "average" feature vector of a client, it disregards the per-round feature distributions unlike \our, which can lead to a lower accuracy when these distributions differ over the rounds (see Fig.~\ref{fig:feature_dist} for illustration). 
The difference between the two approaches becomes significant when $M_r$ is inaccurate and the distributions  $f^+_r$ and $f^-_r$ have a larger overlap (i.e., their difference is smaller towards the end of the training, which indicates decreasing confidence, while $h$ always assigns the same confidence to any feature value across rounds independently of $r$). REG only uses the accuracy of $M_r$ as weights $\mathbf{v}$ in linear regression,  while \our considers, in addition, all feature distributions directly during optimization, which is more accurate.
To confirm this, we report the overlapping coefficient (OVL) between these distributions in Figure \ref{fig:OVL}.b, which 
shows that this coefficient is almost 0 over all rounds when the detection of a gradient inversion attack is considered (on CIFAR-10 with local dataset size $|D_i|=25$). 
This explains why the performances of \our and REG are almost the same (Figure~\ref{fig:inv_attack}.g). However, OVL increases over the rounds for MIA (Figure~\ref{fig:OVL}.a), and \our becomes superior to REG (Figure~\ref{fig:mia_attack}.g).

\subsubsection{Impact of round number} In general, the accuracy of all approaches increases with the number of rounds. This is expected because the number of different observed aggregates also increases, which makes linear regression more accurate. Feature reconstruction techniques converge faster than BASELINE in general (except Figure~\ref{fig:aga_attack}.h as detailed above): for MNIST and FMNIST, \our and REG reach an accuracy of 0.8 by round 80-100 on gradient inversion and ascent detection due to regularization. 
The variance of the reconstructed features is larger
if the number of observed aggregates is too small at the beginning of the training or the attacker model $M_r$ is too inaccurate towards the end of the training for MIA. The effect of this noise is mitigated in \our and REG by regularization, which is not the case for BASELINE and OLS. Moreover,
\textbf{\our and REG need roughly $2N$ rounds to converge}, while BASELINE generally needs more rounds. This is remarkable considering the fact that any approach needs at least $N$ rounds to converge.

\subsubsection{Impact of dataset size}
\textbf{As the dataset size increases, all approaches decline on MIA}, as shown in Figure \ref{fig:mia_attack}.a-h. Indeed, as the model update is computed from the average gradient of all training samples, the impact of a single target sample on the update is smaller if the number of training samples is large. This results in larger variance of the linear features (illustrated in  Figure~\ref{fig:feature_dist}), which is mitigated by regularization in \our and REG.
The detections of gradient inversion and ascent attacks are less impacted by the dataset size (Figure \ref{fig:inv_attack} and \ref{fig:aga_attack}) because these attacks directly manipulate the update vectors.

\vspace*{-0.1cm}

\subsubsection{Impact of the number of clients} In order to evaluate the impact of the total number of participants $N$ on property inference, we perform an experiment on CIFAR-10 considering the MIA and gradient ascent attacks. We set $N$ to 50, 100 and 200 but fix the number of positive clients to 5, and the dataset size is 25. Figure~\ref{fig:impact_N} shows that BASELINE and OLS are the most influenced by the increase in the number of clients. For example, BASELINE reaches an F1-score of 100\% with 50 clients on the gradient ascent attack, and then it decreases to 69\% and then to 21\% with 100 and 200 clients, respectively (see Figure~\ref{fig:impact_N}.d-f). There is a similar decrease for  MIA (see Figure~\ref{fig:impact_N}.a-c). Similarly, OLS decreases from 83\% to 62\% and then to 24\% with 50, 100 and 200 clients, respectively, for the gradient ascent attack. \textbf{Although the F1-score also decreases with PROLIN and REG, this accuracy degradation remains less compared to BASELINE and OLS as the number of clients increases}. The increase in the number of clients seems to widen the gap between the F1-score of PROLIN (which remains better) and REG. 




\section{Conclusion}




We showed that secure aggregation fails to protect client-specific information. We proposed a technique called \our that uses a linear model (due to the linearity of aggregation in federated learning) to extract and disaggregate features for the inference of client-specific property information.

We evaluated our approach on two different tasks: membership inference and the detection of poisoning attacks. In membership inference, the goal is to identify clients whose training data includes a specific target record.
In poisoning detection, the goal is to identify clients that launch untargeted poisoning attacks to degrade the accuracy of the global model. We show that, for both tasks, feature-based reconstruction, and linear models in particular, is surprisingly more accurate than earlier gradient-based reconstruction techniques. Our proposal \our outperforms both the state-of-the-art baseline~\cite{lam2021gradient} and our proposed baselines. In addition, \our has more stable accuracy over rounds, converges faster, and is more robust to more complex scenarios such as when the total number of clients increases or when the attacker model is less accurate.


Our approach is passive and therefore undetectable. Although there are techniques to prevent property reconstruction, those approaches usually introduce trade-offs.
For example, Differential Privacy can be used to avoid detection but at the cost of reducing the accuracy of the common model. Similarly, a client can selectively launch poisoning or use only a subset of all training samples in certain rounds, which can make property reconstruction less accurate. However, these countermeasures also imply less effective attacks or the slower convergence of the common model if the local dataset is small.

\our is not limited to membership inference and misbehaving detection, it can disaggregate the linear features of any supervised detector model. Therefore, there are several avenues of future work that are facilitated by our novel approach, such as contribution scoring, where a score is assigned to each participant measuring, the quality of its contribution to the common federated model, even if secure aggregation is employed. This would also allow to detect free-rider attacks, where a selfish participant benefits from the global model without any useful contribution. There are also stealthier poisoning (backdoor) attacks \cite{XieHCL20,WangSRVASLP20} than gradient inversion and ascent, whose evaluation is left for future work.

Although supervised detector models are generally more accurate than any unsupervised approaches~\cite{unsuperdefense}, they are also restricted to detect only the properties that they are trained for.
However, if malicious clients are adaptive and know what detector model the server uses, they can  evade detection by launching stealthy attacks~\cite{xie2020dba} irregularly over the training. 
The extension of \our to unsupervised misbehaving detection belongs to future work. 

\begin{acks}
This work was partially supported by the Helmholtz Association within the project ``Trustworthy Federated Data Analytics (TFDA)'' (ZT-I-OO1 4)  and ELSA – European Lighthouse on Secure and Safe AI funded by the European Union under grant agreement No. 101070617. Views and opinions expressed are however those of the authors only and do not necessarily reflect those of the European Union or European Commission. Neither the European Union nor the European Commission can be held responsible for them. Support by the European Union project RRF-2.3.1-21-2022-00004 within the framework of the Artificial Intelligence National Laboratory. Funded by the European Union (Grant Agreement Nr. 10109571, SECURED Project). Views and opinions expressed are however those of the author(s) only and do not necessarily reflect those of the European Union or the Health and Digital Executive Agency. Neither the European Union nor the granting authority can be held responsible for them.
\end{acks}

\bibliographystyle{ACM-Reference-Format}
\bibliography{references}

\appendix

\begin{figure*}[h!]
\centering
\includegraphics[scale=0.41]{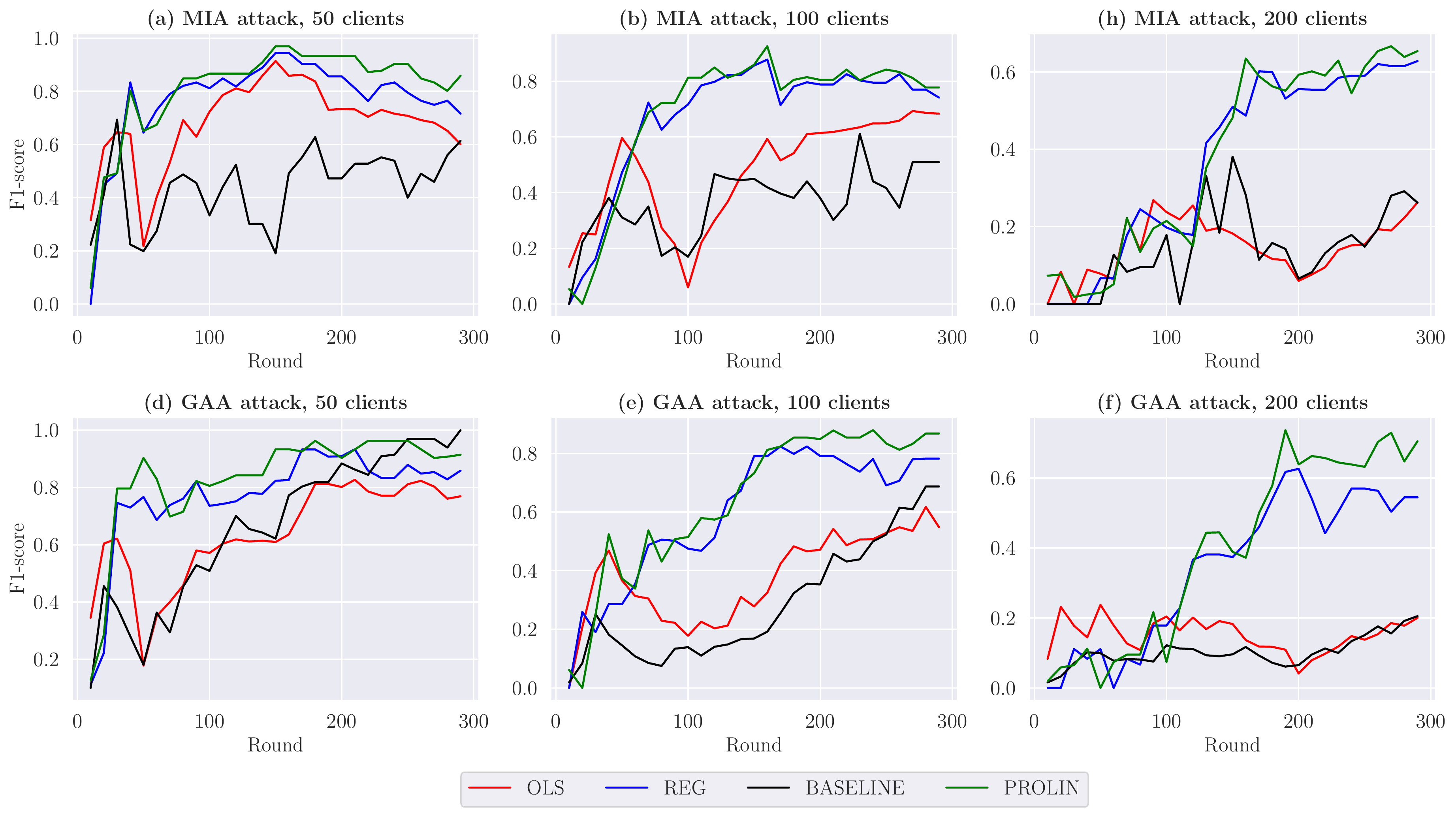}
\caption{Detection of membership inference attacks (MIA) and gradient ascent attacks (GAA) on CIFAR-10 datasets by varying the total number of clients $N$ while keeping the number of clients with the property $P$ to 5.}
\label{fig:impact_N}
\end{figure*}

\section{Analysis}
\label{sec:analysis}
In the following, we provide a theoretical justification of \our. Without loss of generality, suppose that $t=1$. Somewhat abusing the notation, let $\bm{\tau}_i$ denote the probability that client $i$ has property $P$. The maximum likelihood estimation of $\bm{\tau}$ given the observed gradient aggregates $\mathbf{B} = (\mb_1, \ldots, \mb_n)$ is
\begin{align}
\bm{\tau}_{\max} &=  \argmax_{\bm{\tau}} p(\bm{\tau} | \mathbf{B}) \notag \\
&\approx \argmax_{\bm{\tau}} p(\bm{\tau} | \mathbf{G}) \label{eq:ml_approx}
\end{align}
where $p(\cdot|\cdot)$ denotes a generic conditional PDF and $\mathbf{G} = (g_1(\mb_1), \ldots, g_n(\mb_n))$ are the feature aggregates. The last approximation holds if the features extracted by $g_r$ are sufficient to predict the property information, that is, the detector model $M_r = h_r \circ g_r$ is accurate.

Then,
\begin{align}
\bm{\tau}_{\max} &\approx \argmax_{\bm{\tau}} p(\bm{\tau} | \mathbf{G}) \notag \\
&= \argmax_{\bm{\tau}} p(\mathbf{G} | \bm{\tau} )p(\bm{\tau}) \tag{by Bayes' theorem}\\
&= \argmax_{\bm{\tau}}p(\bm{\tau})\int_{\mathbf{X} \in \mathbb{R}
^{n\times N}} \frac{p(\bm{\tau}, \mathbf{G}, \mathbf{X})}{p(\bm{\tau})} d\mathbf{X} \notag \\
&= \argmax_{\bm{\tau}}p(\bm{\tau})\int_{\mathbf{X} \in \mathbb{X}} \frac{p(\bm{\tau}, \mathbf{X})}{p(\bm{\tau})} d\mathbf{X} \notag \\
&= \argmax_{\bm{\tau}}p(\bm{\tau})\int_{\mathbf{X} \in \mathbb{X}} p(\mathbf{X}| \bm{\tau}) d\mathbf{X} 
\end{align}
where  $\mathbb{X} = \{ \mathbf{X}\, |\, \mathbf{X} \in \mathbb{R}^{n\times N} \wedge \forall r:\, \mA_{r,i} \mathbf{X}_{r,i} = \mathbf{G}_r\}$ denotes the set of all individual feature vectors whose per-round aggregates are exactly $\mathbf{G}$.

Assuming a uniform prior on $\bm{\tau}$, we are searching for $\bm{\tau}$ that maximizes $\int_{\mathbf{X}\in \mathbb{X}} p(\mathbf{X} | \bm{\tau})d\mathbf{X}$.
The likelihood function of $\bm{\tau}$ given $\mathbf{X}$ is $L( \bm{\tau} | \mathbf{X}) = p(\mathbf{X} | \bm{\tau})$, 
therefore
\begin{align}
    \argmax_{\bm{\tau}} \int_{\mathbf{X} \in \mathbb{X}} p(\mathbf{X} | \bm{\tau}) d\mathbf{X}&= 
    \argmax_{\bm{\tau}} \log \int_{\mathbf{X} \in \mathbb{X}} L( \bm{\tau} | \mathbf{X}) d\mathbf{X} \label{eq:enum}
\end{align}

Let $\mathcal{Y}_i$ be a Bernoulli random variable, where $\mathcal{Y}_i = 1$ if client $i$ has property $P$.
The likelihood function is 
\begin{align}
    L(\bm{\tau}|\mathbf{X}) &=     p(\mathbf{X}| \bm{\tau})  \notag\\
    &= \prod_i p(\mathbf{X}| \tau_i) \tag{by independence of clients}\\
    &= \prod_{i} \big( p(\mathbf{X}| \mathcal{Y}_i=1, \tau_i) p(\mathcal{Y}_i=1 | \tau_i) \notag \\ 
    & \hspace{1cm}+  p(\mathbf{X}| \mathcal{Y}_i=0, \tau_i) p(\mathcal{Y}_i=0 |\tau_i)  \big) \notag \\
    &= \prod_{i} \left( \tau_i \cdot  p(\mathbf{X}| \mathcal{Y}_i=1, \tau_i)
    + (1-\tau_i) p(\mathbf{X}| \mathcal{Y}_i=0, \tau_i) \right) \label{eq:ml} 
\end{align}

Plugging Eq.~\eqref{eq:ml} into Eq.~\eqref{eq:enum}, we obtain:
\begin{align}
    \bm{\tau}_{\max} = \argmax_{\bm{\tau}} \log \int_{\mathbf{X} \in \mathbb{X}} \prod_{i} \tau_i \cdot  p(\mathbf{X}| \mathcal{Y}_i=1, \tau_i) +\notag \\
     + (1-\tau_i) p(\mathbf{X}| \mathcal{Y}_i=0, \tau_i) \ d\mathbf{X}
     \label{eq:orig}
\end{align}

The exact computation of Eq.~\eqref{eq:orig} is usually infeasible in practice because $p(\mathbf{X}| \mathcal{Y}_i, \tau_i)$ can be specific to client $i$, however, the server may not have any client-specific prior. The best strategy for the server is to approximate  $P(\mathbf{X}| \mathcal{Y}_i, \tau_i)$ with the feature distributions derived from its auxiliary dataset $D^{\mathit{aux}}$. Specifically, 
\begin{align}
p(\mathbf{X}| \mathcal{Y}_i=1, \tau_i) &= \prod_{r \in R(i)} p(\mathbf{X}_{r,i} | \mathbf{X}_{r-1,i}, \ldots, \mathbf{X}_{1,i}, \mathcal{Y}_i=1, \tau_i   ) \notag \\
&\approx \prod_{r\in R(i)} p(\mathbf{X}_{r,i} | \mathcal{Y}_i=1, \tau_i   ) \notag \\
&\approx \prod_{r \in R(i)} f_r^+(\mathbf{X}_{r,i}) \label{eq:approx1} \\ 
p(\mathbf{X}| \mathcal{Y}_i=0, \tau_i) &= \prod_{r \in R(i)} p(\mathbf{X}_{r,i} | \mathbf{X}_{r-1,i}, \ldots, \mathbf{X}_{1,i}, \mathcal{Y}_i=0, \tau_i   ) \notag \\
&\approx \prod_{r \in R(i)} p(\mathbf{X}_{r,i} |  \mathcal{Y}_i=0, \tau_i   ) \notag \\
&\approx \prod_{r \in R(i)} f_r^-(\mathbf{X}_{r,i}) \label{eq:approx2}
\end{align}
for a given client $i$. Here, we assumed that the linear features $\mathbf{X}_{1,i}, \ldots, \mathbf{X}_{n,i}$ of the same client $i$ are independent, which is not true; even if the linear map $g_r$ is different per round, its values are expected to concentrate around $\hat{\mathbf{g}}_i$ for a given $i$ as defined by Eq.~\eqref{eq:lin_model} and approximated by linear regression in Eq.~\eqref{eq:prop_inf_grad}. This explains Constraint 2 in \our.  

The integral in Eq.~\eqref{eq:orig} can be approximated with Monte Carlo integration, which takes several samples from $\mathbb{X}$. These samples should have a large likelihood $L(\bm{\tau}|\mathbf{X})$ since such values are more significant to the integral. Hence, they should be re-sampled if $\bm{\tau}$ changes, which can make the whole optimization process very slow.
To avoid this large overhead of re-sampling, we use the best single sample estimate that maximizes the likelihood. This is a lower bound of the likelihood because    
\begin{align}
\label{eq:ml_deriv_2}
\log \int_{\mathbf{X} \in \mathbb{X}} L( \bm{\tau} | \mathbf{X}) d\mathbf{X} \geq  \log \max_{\mathbf{X} \in \mathbb{X}}  L( \bm{\tau} | \mathbf{X})
\end{align}
Specifically,
\begin{align} 
\argmax_{\bm{\tau}} \int_{\mathbf{X} \in \mathbb{X}} P(\mathbf{X} | \bm{\tau}) d\mathbf{X} \approx 
    \argmax_{\bm{\tau}} \max_{\mathbf{X} \in \mathbb{X}} \log  L( \bm{\tau} | \mathbf{X}) \label{eq:approx}
\end{align}


Therefore, combining Eq.~\eqref{eq:approx}~\eqref{eq:approx1}~\eqref{eq:approx2} with Eq.~\eqref{eq:orig}, we get:
\begin{align*}
    &\bm{\tau}_{\max} \approx \argmax_{\bm{\tau}} \log \int_{\mathbf{X} \in \mathbb{X}} L(\bm{\tau}|\mathbf{X}) d\mathbf{X} \approx \\
    &\approx \argmax_{\bm{\tau}} \max_{\mathbf{X} \in \mathbb{X}}  \sum_{i} \log \Bigg( \tau_i \cdot \prod_r f_r^+(\mathbf{X}_{r,i}) 
    \notag \\ & \hspace{1cm} + (1-\tau_i)\cdot \prod_r f_r^-(\mathbf{X}_{r,i})\Bigg)
\end{align*}
\our performs exactly this optimization with the constraints that $\mathbf{X} \in \mathbb{X}$ (Constraint 1), $\mathbf{X}_{1,i}, \ldots, \mathbf{X}_{n,i}$ concentrate around the solution $\hat{\mathbf{G}}_i$ of Eq.~\eqref{eq:prop_inf_grad} for each client $i$ (Constraint 2), and $\tau_i \in \{0,1\}$ as a client either has or does not have property $P$ (Constraint 3). 

The approximations in Eq.~\eqref{eq:ml_approx} and \eqref{eq:approx} introduce bias into \our, but they also decrease the variance of its prediction compared to gradient reconstruction. If the gradient size $z$ is much larger than the feature size $t$, and $g_r$ effectively extracts all property relevant information, then the variance reduction can outbalance the bias, hence \our can overcome gradient reconstruction. However, if $M_r$ is inaccurate (i.e., $g_r$ cannot capture property relevant information), or $D^{aux}$ is not representative, then the bias can outbalance the variance and gradient reconstruction becomes better.

\section{Comparison of gradient and feature reconstruction}  
\label{sec:comparison}

We compare the reconstruction error of OLS in the gradient space with its error in the feature space.
We show that if property information is scattered across several coordinates of the update vector, then OLS in the feature space has a smaller error than in the gradient space.

Following from Eq.~\eqref{eq:lin_grad}, the linear model is 
$$
\mathbf{B} = \mathbf{A} \mathbf{\hat{W}} + \bm{\Omega}
$$
for gradient aggregation and
$$
\mathbf{G} = \mathbf{A} \mathbf{\hat{G}} + \bm{\Theta}
$$
for feature reconstruction, where $\bm{\Omega}_{i,j} \in \mathbb{R}^{n \times z}$ and  $\bm{\Theta}_{i,j} \in \mathbb{R}^{n \times t}$  are random values that are assumed to have identical normal distributions with variance $\sigma$ just for the sake of comparison.
Assume that OLS is used to obtain an approximation $\mathbf{\widetilde{W}}$ of $\mathbf{\hat{W}}$ and also an approximation $\mathbf{\widetilde{G}}$ of $\mathbf{\hat{G}}$, which means that $\mathbf{\widetilde{W}} = \mathbf{\hat{W}} + \mathbf{A}^+\bm{\Omega}$ and $\mathbf{\widetilde{G}} = \mathbf{\hat{G}} + \mathbf{G}^+\bm{\Theta}$ if $\mathbf{A}$ has full rank
(recall that $\mathbf{A}^+ \in \mathbb{R}^{N\times n}$ is the pseudo-inverse of $\mathbf{A}$). Therefore, 
\begin{align*}
E||g(\mathbf{\hat{W}}) - g(\mathbf{\widetilde{W}})||_1 &= E||g(\mathbf{\hat{W}}) - g(\mathbf{\hat{W}} + \mathbf{A}^+\bm{\Omega})||_1 \\
&= E||g(\mathbf{A}^+\bm{\Omega})||_1 \tag{by linearity of $g$}\\
&= \mathcal{O}(||\bm{\alpha}||_2 N\sqrt{n}\sigma)
\end{align*}
and 
\begin{align*}
E||\mathbf{\hat{G}} - \mathbf{\widetilde{G}}||_1 &=
E||\mathbf{\hat{G}} - (\mathbf{\hat{G}} + \mathbf{A}^+\bm{\Theta})||_1  \\
&= E||\mathbf{A}^+\bm{\Theta}||_1 \\
&= \mathcal{O}(tN\sqrt{n}\sigma)
\end{align*}
where $g(\mathbf{x})= \bm{\alpha} \mathbf{x}$ is fixed over the rounds and $(\mathbf{A}^+_{i,j})^2 = \mathcal{O}(1)$\footnote{Each element of 
$\mathbf{A}^+\bm{\Omega} \in \mathbb{R}^{N\times z}$ is a normal random variable with mean 0 and variance $\sum_{j}(\mathbf{A}^+_{i,j})^2 \sigma^2$,  hence $[g(\mathbf{A}^+\bm{\Omega})]_i$ is also a normal random variable with mean 0 and variance $\sum_{k=1}^z\bm{\alpha}_{k}^2\sum_{j=1}^n(\mathbf{A}^+_{i,j})^2 \sigma^2$.}.

Therefore, the error when gradients are reconstructed is larger with a factor of $\mathcal{O}(||\bm{\alpha}||_2/t)$.
Since $\bm{\alpha}_i$ represents the impact of gradient coordinate $i$ on property inference, gradient- and feature-based reconstructions can have comparable performance if $\bm{\alpha}$ is sparse (i.e., when the property information is already encoded by only a few gradient coordinate values whose reconstructions are sufficient for successful inference).

\section{More details}
All settings are summarized in Table \ref{tab:params}.

\begin{table}[h!]
    \centering
    \resizebox{0.7\columnwidth}{!}{%
    \begin{tabular}{llll}
        \toprule
        Notation &  MNIST & Fashion-MNIST & CIFAR-10\\
        \midrule
         $n$ & 300 & 300 & 300 \\
         $z$ &  21,840  & 21,840 & 62,006 \\
         $N$ & 50 & 50 & 50     \\
         $\lambda$ & 5 & 5 & 5 \\
         $C$ & 0.2 & 0.2 & 0.2 \\
         $\phi$ & 0.1 & 0.1 & 0.1 \\
         $\eta$ & 0.01 & 0.01 & 0.1\\
         $|D^{aux}|$ & 6,000 & 6,000 & 5,000 \\
         $|D'|$ & $2|D^{aux}|$ & $2|D^{aux}|$ & $2|D^{aux}|$ \\
        \bottomrule
    \end{tabular}%
}

    \caption{Values of Hyperparameters    \label{tab:params}}
\end{table}

\end{document}